\documentclass{emulateapj}

\usepackage{graphicx}
\usepackage[]{natbib}
\usepackage{placeins}
\usepackage{lineno}
\usepackage{ulem}
\usepackage{rotating}
\usepackage{adjustbox}

\usepackage[caption=false]{subfig}    

\pdfoutput=1

\newcommand{\degree}{\ensuremath{^\circ}}

\newcommand{\ltsima} {$\; \buildrel < \over \sim \;$}

\newcommand{\gtsima} {$\; \buildrel > \over \sim \;$}
\newcommand{\lta} {\lower.5ex\hbox{\ltsima}}
\newcommand{\gta} {\lower.5ex\hbox{\gtsima}}

\shorttitle{GRB~131014A: a Laboratory to Study the Thermal-Like and Non-Thermal Emissions in GRBs}
\shortauthors{S.Guiriec}

\begin{document}

\title{GRB~131014A: a Laboratory to Study the Thermal-Like and Non-Thermal Emissions in Gamma-Ray Bursts, and the new L$_\mathrm{\lowercase{i}}^\mathrm{\lowercase{n}T\lowercase{h}}$--E$_\mathrm{\lowercase{peak,i}}^\mathrm{\lowercase{n}T\lowercase{h,rest}}$ relation}

\author{S. Guiriec\altaffilmark{1,2,3}, R. Mochkovitch\altaffilmark{4}, T. Piran\altaffilmark{5}, F. Daigne\altaffilmark{4}, C. Kouveliotou\altaffilmark{6}, J. Racusin\altaffilmark{1}, N. Gehrels\altaffilmark{1}, and J. McEnery\altaffilmark{1}}




\altaffiltext{1}{NASA Goddard Space Flight Center, Greenbelt, MD 20771, USA}
\altaffiltext{2}{Department of Physics and Department of Astronomy, University of Maryland, College Park, MD 20742, USA}
\altaffiltext{3}{Center for Research and Exploration in Space Science and Technology (CRESST)}
\altaffiltext{4}{Institut d'Astrophysique de Paris â UMR 7095 Universit\'e Pierre et Marie Curie-Paris 06; CNRS 98 bis bd Arago, 75014 Paris, France}

\altaffiltext{5}{Racah Institute of Physics, The Hebrew University of Jerusalem, Jerusalem 91904, Israel 0000-0002-7964-5420}
\altaffiltext{6}{Department of Physics, The George Washington University, Washington, DC 20052}



\email{sylvain.guiriec@nasa.gov}

\begin{abstract}

Over the past years, evidence has been accumulated on the existence of a thermal-like component during the prompt phase of Gamma Ray Bursts (GRBs). However, this component, often associated with the GRB jet's photosphere, is usually subdominant compared to a much stronger non-thermal one. The prompt emission of GRB~131014A---detected by the {\it Fermi Gamma-ray Space Telescope} (hereafter {\it Fermi})---provides an unique opportunity to trace the history of this thermal-like component. Indeed, the thermal emission in GRB~131014A is much more intense than in other GRBs and a pure thermal episode is observed during the initial 0.16 s. The thermal-like component cools monotonically during the first second while the non-thermal emission kicks off. The intensity of the non-thermal component progressively increases until being energetically dominant at late time similarly to what is typically observed. This is a perfect scenario to disentangle the thermal component from the non-thermal one. The initial decaying and cooling phase of the thermal-like component is followed by a strong re-brightening and a re-heating episode; however, despite a much brighter second emission phase, the temperature of the thermal component does not reach its initial value. This re-brightening episode is followed by a global constant cooling until the end of the burst. We note that a shallower low-energy spectral slope than the typical index value +1, corresponding to a pure Planck function, better matches with the thermal-like spectral shape; a spectral index around +0.6 seems to be in better agreement with the data. The non-thermal component is adequately fitted with a Band function whose low and high energy power law indices are $\sim$-0.7 and $<\sim$-3, respectively; this is also statistically globally equivalent to a cutoff power law with a $\sim$-0.7 index. This is in agreement with our previous results. Finally, a strong correlation is observed between the time-resolved energy flux, F$_\mathrm{i}^\mathrm{nTh}$, and the corresponding spectral peak energy, E$_\mathrm{peak,i}^\mathrm{nTh}$, of the non-thermal component with a slope similar to the one reported in our previous articles. Assuming a universal relation between the time-resolved luminosity of the non-thermal component, L$_\mathrm{i}^\mathrm{nTh}$, and its rest frame E$_\mathrm{peak,i}^\mathrm{nTh}$, E$_\mathrm{peak,i}^\mathrm{rest,nTh}$, that we derived from a limited sample of GRBs detected by {\it Fermi}, we estimate a redshift of $\sim$1.55 for GRB~131014A that is a typical value for long GRBs.
These observational results are consistent with the models in which the non-thermal emission is produced well above the GRB jet photosphere but they may also be compatible with other scenarios (e.g., dissipative photosphere) that are not discussed in this article.

\end{abstract}

\keywords{Gamma-ray burst: individual: GRB~131014A  -- Gamma-ray burst: individual: GRB~131014A  -- Radiation mechanisms: thermal -- Radiation mechanisms: non-thermal -- Acceleration of particles}

\section{Introduction}

\begin{figure*}[ht!]
\begin{center}
\includegraphics[totalheight=0.54\textheight, clip]{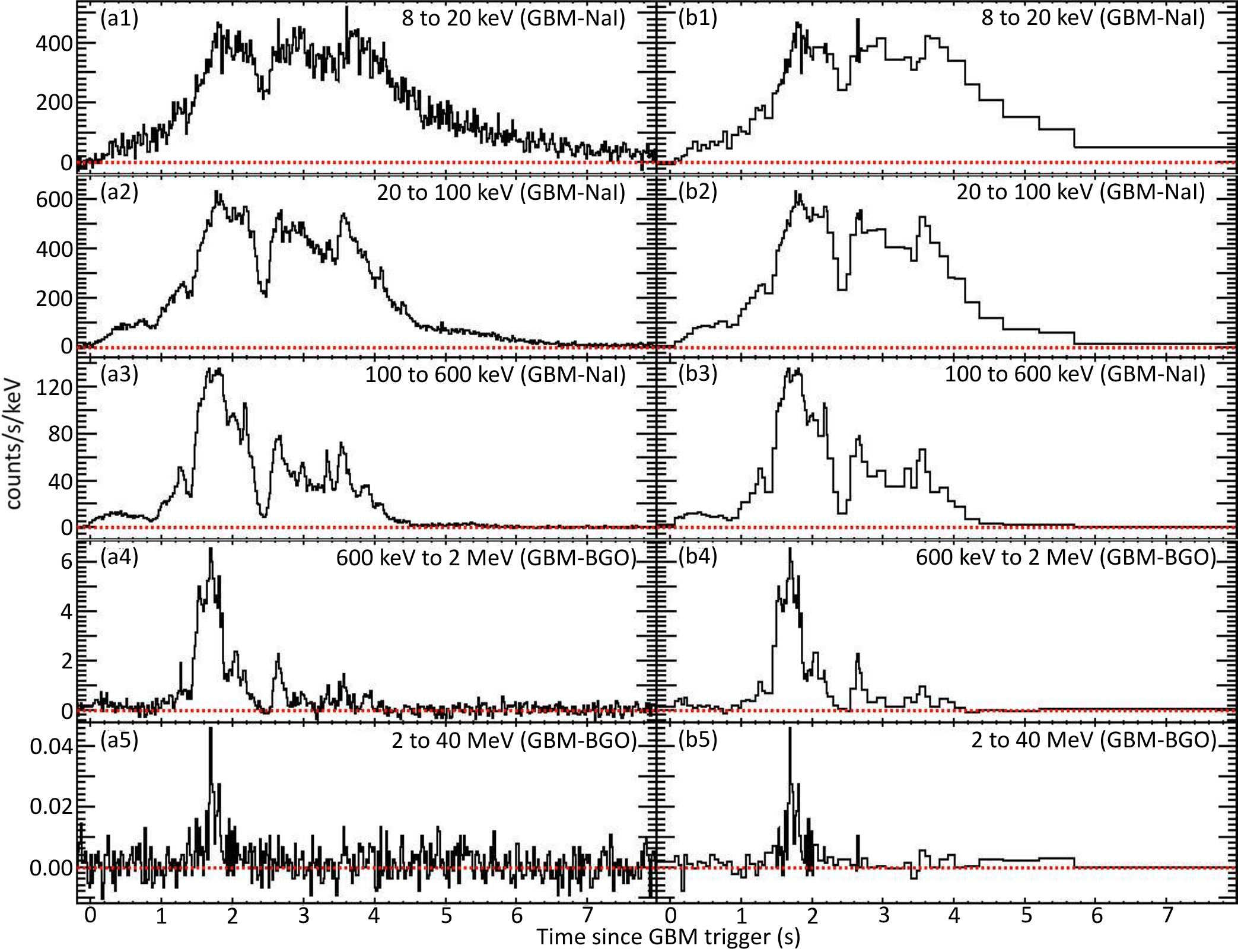}
\caption{\label{fig01}Count light curves as observed with GBM NaI (8--600 keV) and BGO (600 keV--40 MeV) detectors. (a1--5) 0.02 s time-resolved light curves. (b1--5) Light curves with the binning used for the time-resolved spectral analysis.}
\end{center}
\end{figure*}


During the past few years, the paradigm for the prompt emission of both short and long Gamma Ray Bursts (GRBs) has evolved from a single-hump keV--MeV $\nu$F$_\nu$ spectrum to a double-hump scenario~\citep{Guiriec:2011,Guiriec:2013a,Guiriec:2015}; it has been proposed that the two humps correspond to two separate emission processes: (i) a thermal-like component---usually adequately approximated with a black body (BB) spectrum---that may be the footprint of the GRB jet photosphere predicted by the fireball model~\citep{Cavallo:1978,Paczynski:1986,Goodman:1986,Shemi:1990,Rees:1992,Meszaros:1993,Rees:1994} and (ii) a non-thermal like component that may be related to particle acceleration processes within the GRB jet such as internal shocks~\citep{Rees:1994,Kobayashi:1997,Daigne:1998} and/or magnetic reconnection mechanisms~\citep{Zhang:2011b,Zhang:2014}. Despite its non-thermal shape, this component may also be interpreted as strongly reprocessed photospheric emission episodes~\citep{Beloborodov:2010,Beloborodov:2014,Vurm:2015}.

The so-called empirical Band function~\citep[hereafter Band for shorter --][]{Band:1993,Greiner:1995} with its single energy spectral break has been traditionally considered as a good description of the prompt emission spectra when fitted alone to the data. Band is a smoothly broken power-law (PL) defined with four parameters: the indices $\alpha_\mathrm{Band}$ and $\beta_\mathrm{Band}$ of the low and high energy PLs, respectively, the energy E$_\mathrm{peak}^\mathrm{Band}$ at the maximum of the $\nu$F$_\nu$ spectrum \citep{Gehrels:1997}, and a normalization parameter. Despite the typical non-thermal shape of Band \citep[see for instance the results of the GBM spectral catalog--][]{Gruber:2014}, it is often incompatible with the synchrotron emission scenarios that are usually considered as the most likely ones to account for GRB prompt emission. Indeed, the values of $\alpha_\mathrm{Band}$ are often $>$-3/2 and even $>$-1/3, the limits of the synchrotron emission from electrons in the pure slow and fast cooling regimes, respectively \citep{Cohen:1997,Crider:1997,Ghisellini:2000}.

When Band is fitted to the data together with a BB component (i.e., Band+BB), the values of $\alpha_\mathrm{Band}$ are much more in agreement with the synchrotron model predictions compared to the values of $\alpha_\mathrm{Band}$ resulting from the fit to Band-alone \citep{Guiriec:2011,Guiriec:2013a,Guiriec:2015}. In the Band+BB scenario, the thermal-like component is usually subdominant compared to the non-thermal one that is consistent with the theoretical predictions of~\citet{Daigne:2002}, ~\citet{Nakar:2005}, ~\citet{Zhang:2011b}, and ~\citet{Hascoet:2013}. Moreover, in the Band+BB scenario, Band can often be replaced with a power law with an exponential cutoff (CPL) because of the steep spectral slope of Band at high energies (i.e., $\beta_\mathrm{Band}$$\lesssim$-3.)

In \citet{Guiriec:2015} we showed some evidence that GRB prompt emission spectra---observed with the {\it Fermi Gamma-ray Space Telescope} (hereafter {\it Fermi})---are adequately described with a combination of three separate components, C$_\mathrm{nTh}$+C$_\mathrm{Th}$+PL: (i) a non-thermal component (C$_\mathrm{nTh}$), (ii) a thermal-like component (C$_\mathrm{Th}$) and (iii) an additional PL with or without a cutoff.  The three components are not systematically present in all bursts, especially the additional PL. We also concluded that in the context of this multi-component model, the values of low-energy PL spectral index, $\alpha_\mathrm{nTh}$, do not evolve much during the prompt phase and remain constant around either -0.7 or -1.2 depending on the burst. \citet{Guiriec:2015c} confirmed these results using GRBs observed with the Burst And Transient Source Experiment (BATSE) on board the {\it Compton Gamma-Ray Observatory} (CGRO).

As a result of this new multi-component model, a strong correlation has been found between the time-resolved energy flux of C$_\mathrm{nTh}$, F$_\mathrm{i}^\mathrm{nTh}$, and its corresponding $\nu$F$_\nu$ peak energy, E$_\mathrm{peak,i}^\mathrm{nTh}$ \citep[hereafter F$^\mathrm{nTh}_\mathrm{i}$--E$_\mathrm{peak,i}^\mathrm{nTh}$ relation;][]{Guiriec:2013a,Guiriec:2015}\footnote{The index ``i'' refers to the instantaneous spectra within a single GRB.}. When fitted to a PL, the indices of the F$^\mathrm{nTh}_\mathrm{i}$--E$^\mathrm{nTh}_\mathrm{peak,i}$ relations are very similar for all GRBs (i.e., $\sim$1.4) and when corrected for the redshift (i.e., in the source frame) a universal relation for all GRBs appears between the time-resolved luminosities of the non-thermal component, L$^\mathrm{nTh}_\mathrm{i}$ and its $\nu$F$_\nu$ peak energy in the rest frame, E$_\mathrm{peak,i}^\mathrm{nTh,rest}$ \citep[hereafter L$^\mathrm{nTh}_\mathrm{i}$--E$_\mathrm{peak,i}^\mathrm{nTh,rest}$ --][]{Guiriec:2013a,Guiriec:2015}.

Because C$_\mathrm{Th}$ is usually energetically subdominant, it is often challenging to capture its exact spectral shape. We showed in \citet{Guiriec:2011,Guiriec:2013a,Guiriec:2015} that this component must be thermal by replacing it with a CPL that required a positive index; however, a pure BB component was as good an approximation overall and this option was chosen to limit the number of free parameters. GRB~131014A---observed with the {\it Fermi}---is a very bright burst exhibiting an unusually strong C$_\mathrm{Th}$ component that accounts for approximately half of the total energy in the prompt emission. In this burst only C$_\mathrm{Th}$ is detected during the very beginning of the prompt phase before the appearance of C$_\mathrm{nTh}$. For these reasons, GRB~131014A is a perfect laboratory for the study of the spectral shape and behavior of C$_\mathrm{Th}$ as well as the relative evolution and correlation of C$_\mathrm{Th}$ and C$_\mathrm{nTh}$.

In Sections \ref{section:observation}, \ref{section:dataAnalysis} and \ref{section:method} we present the observations, the data selection and the analysis methodology, respectively. In Section \ref{sec:results}, we report the results of our analysis: (i) the initial thermal emission episode, (ii) the description of the C$_\mathrm{nTh}$+C$_\mathrm{Th}$ scenario, (iii) the comparison of C$_\mathrm{nTh}$+C$_\mathrm{Th}$ with the Band-alone fit results, and (iv) the redshift estimates using the new L$_\mathrm{i}^\mathrm{nTh}$--E$_\mathrm{peak,i}^\mathrm{nTh,rest}$ relation. Finally, in Section \ref{section:interpretation} we briefly examine how these new observational results support the models in which the non-thermal emission is produced well above the photosphere.

\vspace{-0.2cm}
\section{Observations}
\label{section:observation}
The {\it Fermi}/Gamma-ray Burst Monitor (GBM) detected GRB~131014A on 2013 October 14 at T$_\mathrm{0}$=05:09:00 UT~\citep{{Fitzpatrick:2013}}. Its T$_\mathrm{90}$ duration between 50 and 300 keV~\citep{Kouveliotou:1993} is $\sim$3.2 s, which puts it in the class of long GRBs (i.e., T$_\mathrm{90}$$>$2 s). The burst was also detected with the {\it Fermi}/Large Area Telescope (LAT) at high energies with the highest energy photon observed at 1.8 GeV about 15 s after the GBM trigger time~\citep{Desiante:2013}. The X-Ray Telescope (XRT) on-board {\it Swift} located the GRB at RA = 100.303\degree~and Dec = -19.097\degree ($\pm$0.001\degree), which was $\sim$74\degree from the LAT boresight at the burst trigger time \citep{Evans:2013}. We show the GBM light curves of  GRB~131014A in several energy ranges in Figure~\ref{fig01}.

\section{Data Selection}
\label{section:dataAnalysis}
GBM consists of 12 Sodium Iodide (NaI) detectors covering the energy range from 8 keV to 1 MeV. For this analysis, we selected the optimal detectors with an angle to the source $<$30$\degree$ based on the best source location above and with no blockage by another part of the spacecraft (e.g., LAT, radiators, solar panels) as well as with no shadowing by another GBM module. According to these criteria, we selected detectors n9, na and nb\footnote{The NaIs are named nx with x varying from 0 to 9 for the first 10 detectors, and ``a'' and ``b'' for detectors 11 and 12, respectively.}.

GBM also includes two bismuth germanate (BGO) detectors (b0, b1) covering the energy range from 200 keV to 40 MeV and located on the opposite sides of the spacecraft. Usually, the source is only in the field of view of one BGO detector, so we retained b1 for GRB~131014A.

Here, we analyzed GBM time tagged event (TTE) data that have the finest time (i.e., 2 $\mu$s) and energy resolution that can be achieved with GBM. We used the NaI data from 8 keV to $\sim$900 keV cutting out the overflow high energy channels as well as the K-edge from $\sim$30 to $\sim$40 keV. We also used the BGO data from 200 keV to $\sim$40 MeV cutting out the overflow energy channels. We then generated the response files for each GBM detector based on the best source location.

The background in each of the GBM detectors was estimated by fitting polynomial functions to the light curves in various detector energy channels before and after the source active time period. The background was then estimated by interpolating the functions to the source active time period.

Detailed information about the GBM can be found in~\citet{Meegan:2009}\footnote{See also http://fermi.gsfc.nasa.gov for additional and up to date information about \it{Fermi}.}.

\section{Analysis Methodology}
\label{section:method}
We performed a fine-time analysis of GRB~131014A following the same procedure as presented in~\citet{Guiriec:2010,Guiriec:2011,Guiriec:2013a,Guiriec:2015}. The time intervals used for this analysis are displayed in Figure~\ref{fig01}b. Similarly to our previous works we fitted to each time interval various models that we can summarize as follows: C$_\mathrm{nTh}$+C$_\mathrm{Th}$+C$_\mathrm{nTh2}$ where (i) C$_\mathrm{nTh}$ is the traditional keV-MeV non-thermal component that we approximate with Band, CPL or CPL with an index fixed at -0.7 (i.e., CPL$_\mathrm{-0.7}$), (ii) C$_\mathrm{Th}$ is a thermal-like component that we approximate with BB, CPL with an index of +0.6 (CPL$_\mathrm{+0.6}$)\footnote{CPL$_\mathrm{+0.6}$ is broader than a pure Plank function---whose low-energy spectral index is +1---and, therefore, more compatible with the jet photospheric models, which predict a spectral broadening because of the jet curvature and sub-photospheric energy dissipation.}, CPL or Band for more freedom in the spectral shape, and (iii) C$_\mathrm{nTh2}$ is a second non-thermal component extending from a few keV up to hundreds of MeV that we approximate with PL, CPL, PL with a fixed index at -1.5 or a CPL with a fixed index at -1. For direct comparison with what is usually done in the literature we also fitted Band-alone to the data. To perform the spectral analysis, we used the {\it Rmfit} fitting package. The best spectral parameter values resulting from the spectral fits as well as their 1--$\sigma$ uncertainties were estimated by optimizing the Castor C-statistic (hereafter Cstat), which is a likelihood technique converging to a $\chi^2$ for a specific data set when there are enough counts.

\begin{figure*}[ht!]
\begin{center}
\includegraphics[totalheight=0.50\textheight, clip]{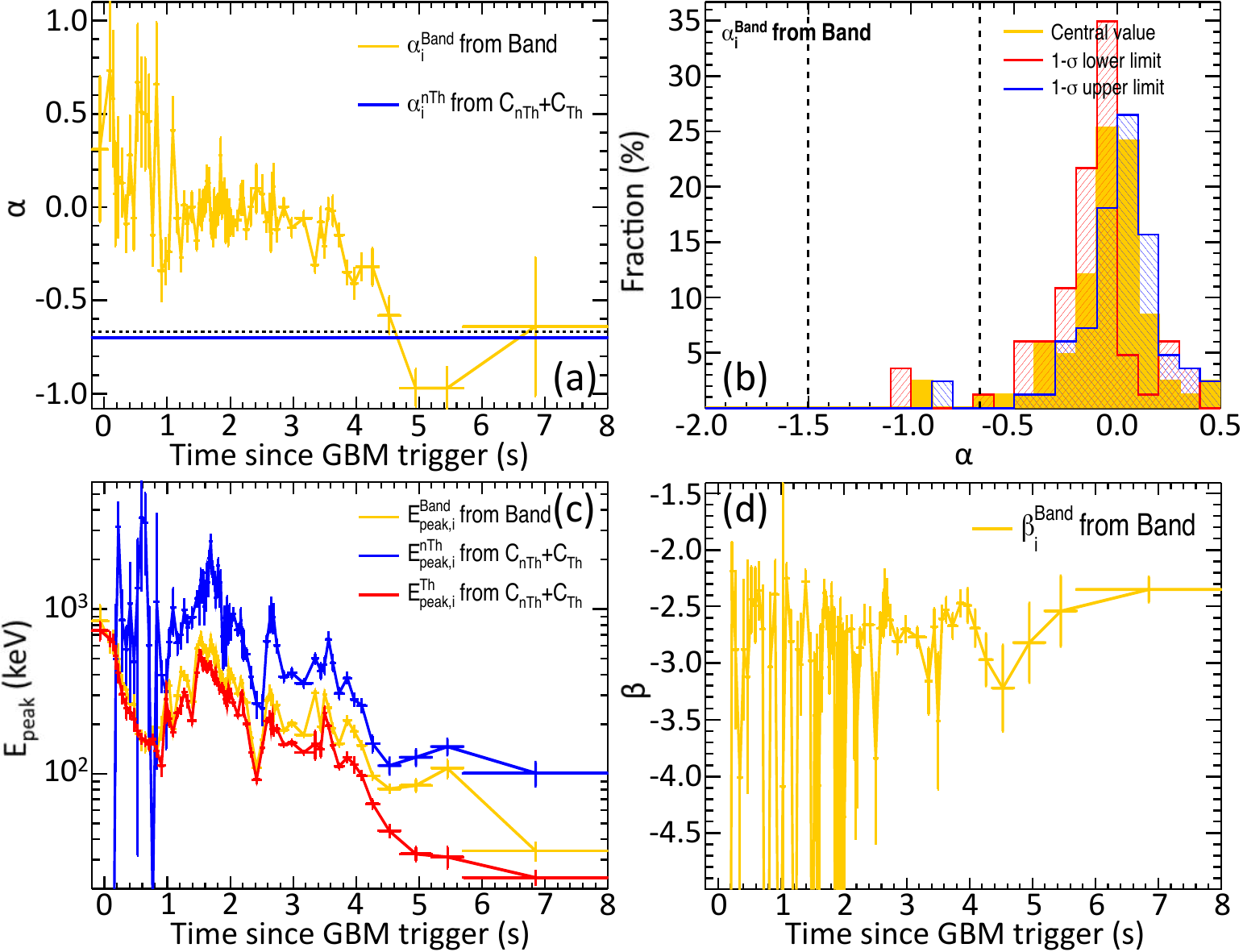}
\caption{\label{fig02}Evolution with time of the spectral parameters resulting from the fits of a Band function alone and of the C$_\mathrm{nTh}$+C$_\mathrm{Th}$ model to the fine time intervals. (a) Evolution of $\alpha$ for Band (orange). For C$_\mathrm{nTh}$+C$_\mathrm{Th}$ (blue), $\alpha$ is fixed to -0.7.  The dashed regions $>$-2/3 and $>$-3/2 correspond to the domains in which the values of $\alpha$ are incompatible with pure synchrotron emission from electrons in both the slow and fast cooling regimes, and with synchrotron emission from electrons in the fast cooling regime only, respectively. (b) Distribution of the values of $\alpha_\mathrm{Band}$ resulting from the Band-only fits. The orange, red and blue histograms correspond to the distributions of the mean, the 1--$\sigma$ lower limit and the  the 1--$\sigma$ upper limit of $\alpha_\mathrm{Band}$, respectively. The vertical dashed lines correspond to the pure synchrotron limits as described in the caption of the top right panel. (c) Evolution of the $\nu$F$_\nu$ spectral peaks of the Band function, E$_\mathrm{peak}^\mathrm{Band}$ (orange), of the non-thermal component C$_\mathrm{nTh}$ of the C$_\mathrm{nTh}$+C$_\mathrm{Th}$ model, E$_\mathrm{peak}^\mathrm{nTh}$ (blue) and of the thermal-like component C$_\mathrm{Th}$ of the C$_\mathrm{nTh}$+C$_\mathrm{Th}$ model, E$_\mathrm{peak}^\mathrm{Th}$ (red). The temperature, kT, of the thermal-like component C$_\mathrm{Th}$ is obtained by dividing E$_\mathrm{peak}^\mathrm{Th}$ by $\sim$2.5. (d) Evolution of Band's $\beta$.}
\end{center}
\end{figure*}

\newpage
\section{Results}
\label{sec:results}

\begin{figure*}[ht!]
\begin{center}
\includegraphics[totalheight=0.22\textheight, clip]{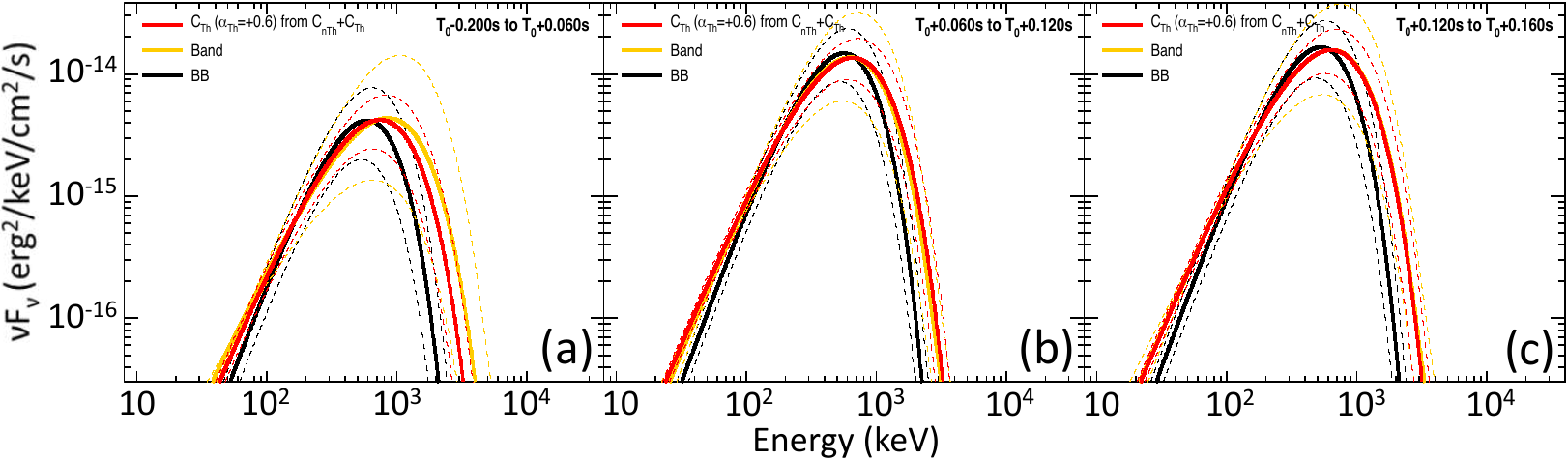}
\caption{\label{fig03}Results of the spectral fits in three time intervals during the initial pure thermal episode using BB (black), Band (orange) and C$_\mathrm{nTh}$+C$_\mathrm{Th}$ (red) with indices fixed at -0.7 and +0.6 for C$_\mathrm{nTh}$ and C$_\mathrm{Th}$, respectively. The non-thermal C$_\mathrm{nTh}$ component of C$_\mathrm{nTh}$+C$_\mathrm{Th}$ does not contribute to the emission in these three time intervals. The dashed lines correspond to the fit 1--$\sigma$ confidence regions.}
\end{center}
\end{figure*}

The most relevant spectral fits are reported in Table~\ref{tab01}. It was not possible to identify any clear  contribution from C$_\mathrm{nTh2}$ in GRB 131014A spectra---but its existence cannot be completely discarded either; therefore, we only considered the two component model, C$_\mathrm{nTh}$+C$_\mathrm{Th}$, and not the three component scenario, C$_\mathrm{nTh}$+C$_\mathrm{Th}$+C$_\mathrm{nTh2}$.

The CPL+BB fits result in slightly lower Cstat values than the Band-alone ones in $\sim$64\% of the time intervals but the former model has one more parameter; however, Cstat improvements up to 21 units can be measured in some time intervals\footnote{In the case of nested models and in the Gaussian regime an improvement of $\sim$10 units of Cstat per additional degree of freedom corresponds to a $\sim$5 $\sigma$ level improvement.}. In the remaining $\sim$36\% of the cases, CPL+BB and Band-alone fits lead either to the same Cstat values or Band is better than CPL+BB for a few units of Cstat\footnote{This is possible because Band and CPL+BB are not nested models.}. In the two brightest time intervals (i.e., from T$_\mathrm{0}$+1.68 s to T$_\mathrm{0}$+1.72 s), Band+BB is better than CPL+BB by $\sim$20 units of Cstat. This can be easily explained by the following argument. Because the limited number of observed counts at high energy in the selected fine time-intervals and because of the intrinsically fast decrease of the flux with energy at the high end of the spectrum at the source (i.e., compatible with a PL with an index $\lesssim$-3\footnote{This has been verified with simulations.}), there is, usually, no difference in fitting the high-energy spectrum either to a PL or to an exponential cut off from statistical considerations; here, the flux in these two time intervals is high enough to discriminate between the two spectral shapes and, therefore, the fit to Band+BB is significantly better than the fit to CPL+BB, and Band+BB results in a high-energy PL index such as -3.12$<$$\beta_\mathrm{Band}$$<$-2.26.


CPL+CPL$_\mathrm{+0.6}$ slightly better fits the data than CPL+BB in all the time intervals; the CPL$_\mathrm{+0.6}$ thermal-like component is consistent with a spectral broadening of the Plank function as predicted by the GRB jet photospheric models. Also, CPL+CPL$_\mathrm{+0.6}$ systematically results in better fits than Band-alone in all time intervals but two (i.e., from T$_\mathrm{0}$+1.68 s to T$_\mathrm{0}$+1.72 s). 

\citet{Guiriec:2015} suggested that the value of the spectral index of the low-energy PL of C$_\mathrm{nTh}$, $\alpha_\mathrm{nTh}$, was either $\sim$-0.7 or $\sim$-1.2. The estimate of $\alpha_\mathrm{nTh}$  when fitting C$_\mathrm{nTh}$+C$_\mathrm{Th}$ to the data is challenging in the case of GRB 131014A because of the large uncertainties on this parameter values; however, the parameter minimization clearly prefers options with $\alpha_\mathrm{nTh}$$\geq$-0.7. Overall, the CPL$_\mathrm{-0.7}$+CPL$_\mathrm{+0.6}$ fits result in similar Cstat values as Band-alone with the same number of free parameters (i.e., 4).
It must be noted that values of $\alpha_\mathrm{nTh}$ slightly larger than -0.7 (i.e., -0.5$\geq$$\alpha_\mathrm{nTh}$$\geq$-0.7) would better fit the data, but without a good estimate we choose to use the value of -0.7 proposed in~\citet{Guiriec:2015}. From here forward, we refer to the CPL$_\mathrm{-0.7}$+CPL$_\mathrm{+0.6}$ scenario as C$_\mathrm{nTh}$+C$_\mathrm{Th}$.


In the following, we will mostly compare the Band-alone and the C$_\mathrm{nTh}$+C$_\mathrm{Th}$ scenarios. The two model spectral shapes in the various time intervals are overplotted in Figure~\ref{fig10}.

In order to better compare the spectral shapes of C$_\mathrm{Th}$ and C$_\mathrm{nTh}$ as well as of the Band function, we plotted in Figure~\ref{fig02}c 
the energy at the maximum of the $\nu$F$_\nu$ spectrum for each component: E$_\mathrm{peak}^\mathrm{Th}$, E$_\mathrm{peak}^\mathrm{nTh}$ and E$_\mathrm{peak}^\mathrm{Band}$ for C$_\mathrm{Th}$, C$_\mathrm{nTh}$ and Band, respectively. The actual temperature, kT, of the thermal-like component, C$_\mathrm{Th}$, is reported in Table~\ref{tab01}. 

\vspace{-0.05cm}
\subsection{An Initial 0.16 s Pure Thermal Episode}

Over the whole burst duration, the values of $\alpha_\mathrm{Band}$ resulting from the Band-only fits are unusually high with a cluster around 0 (see Figure~\ref{fig02}b). This is particularly striking during the first 1 second of the burst where the $\alpha_\mathrm{Band}$ values are between 0 and 1 with a mean around +0.5 (see Figure~\ref{fig02}a). From $\sim$1 s to $\sim$3.6 s, the $\alpha_\mathrm{Band}$ values remain constant around 0 before dropping down to -1 thereafter. Those very high $\alpha_\mathrm{Band}$ values rule out non-thermal emission processes alone to account for the observations.

The first 1 second of the burst corresponds to a low intensity plateau visible in the light curves of Figure~\ref{fig01} from 8 to 600 keV. Between 600 keV and 2 MeV the plateau is not seen but a little excess around the trigger time precedes the main emission
starting one second later.

During the first 0.16 s, GRB~131014A time-resolved spectra have a thermal-like shape. A pure BB is a good description of the spectra and no additional non-thermal component can be identified contrary to later time intervals similarly to the results reported in~\citet{Ghirlanda:2003},~\citet{Ryde:2004} and~\citet{Guiriec:2015c} for GRB~970111 observed with the Burst And Transient Source Experiment (BATSE) on board the Compton Gamma-Ray Observatory (CGRO). However, in using models with more freedom for the spectral shape than BB, such as Band or CPL, the fitting engine prefers solutions with broader curvatures than the pure BB one. Indeed, the fits to a Band function or to a CPL component alone result in spectral indices, for the low-energy PL, between 0 and +1. A CPL with an index fixed at +0.6---which is more consistent with the jet's photosphere models that predict broader curvature---has the same number of parameters as a pure BB and slightly better fits the data on the overall GRB duration. This is illustrated in Figure~\ref{fig03} where the fit results using BB, Band-only and CPL with a +0.6 index are overplotted; while C with a +0.6 index overlaps very well with Band, whose shape has much more freedom to fit the data, BB is clearly too narrow.

While the thermal-component is usually observed together with the non-thermal one \citep{Guiriec:2011,Guiriec:2013a,Guiriec:2015}, GRB~131014A is the first case exhibiting a pure initial thermal episode before the non-thermal emission kicks off.
In addition, GRB~131014A shows that the thermal emission may start prior to the non-thermal one.

\begin{figure*}[ht!]
\begin{center}
\includegraphics[totalheight=0.59\textheight, clip]{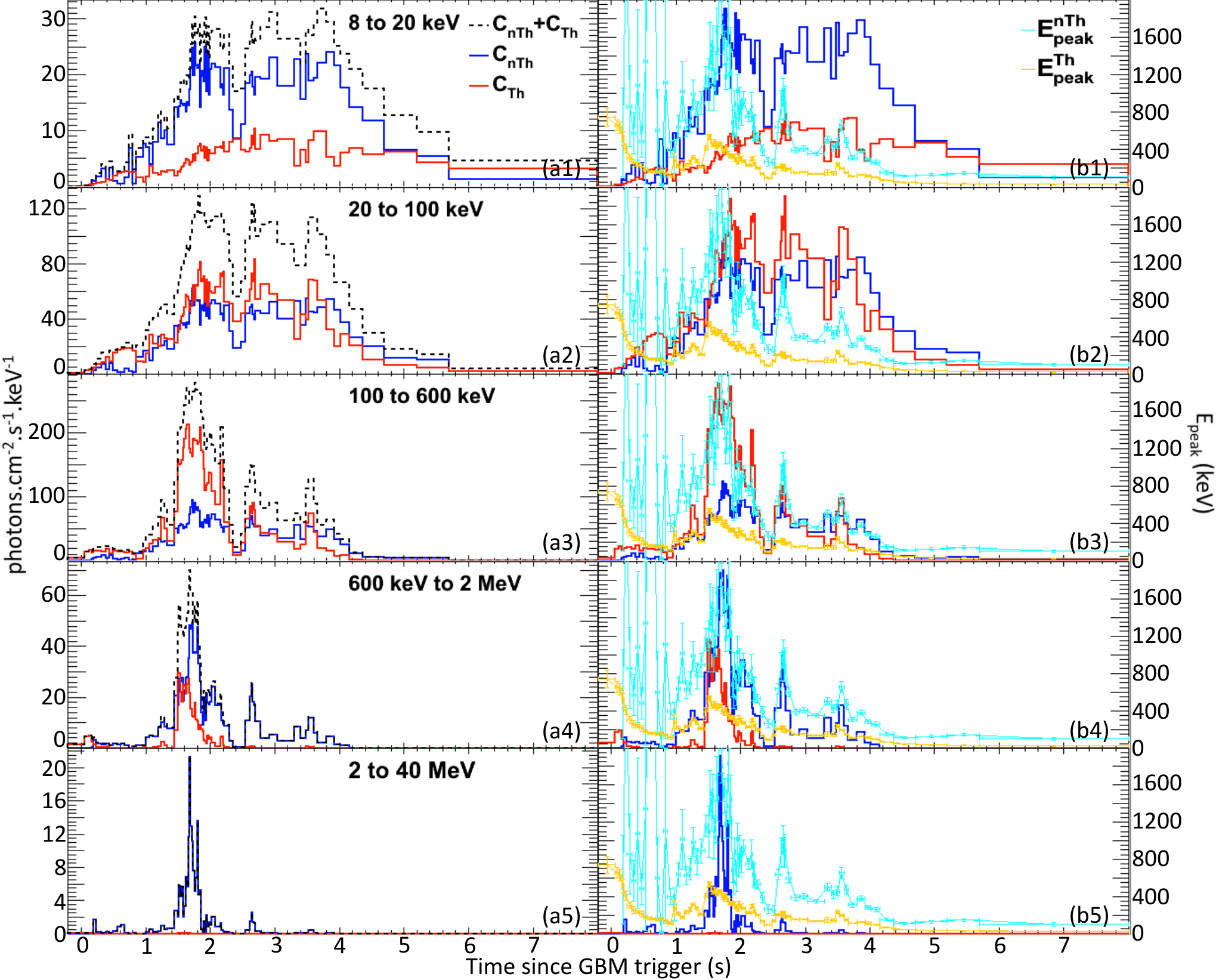}
\caption{\label{fig04}Reconstructed photon light curves resulting from the spectral analysis using the C$_\mathrm{nTh}$+C$_\mathrm{Th}$ (with $\alpha_\mathrm{nTh}$=-0.7 and $\alpha_\mathrm{Th}$=+0.6) model. The light curves are displayed in the same energy bands as the count light curves presented in Figure~\ref{fig01} and with the same time intervals as in Figure~\ref{fig01}b. (a1--5) The reconstructed photon light curves of the non-thermal (C$_\mathrm{nTh}$) and of the thermal-like (C$_\mathrm{Th}$) components are displayed in blue and red, respectively. The black dashed lines correspond to the sum of the two component (C$_\mathrm{nTh}$+C$_\mathrm{Th}$). (b1--5) The non-thermal and thermal-like component are displayed in blue and red, respectively, together with the evolution of the $\nu$F$_\nu$ spectral peaks of the non-thermal component, E$_\mathrm{peak}^\mathrm{nTh}$, in cyan and of the thermal-like component, E$_\mathrm{peak}^\mathrm{Th}$, in orange. The temperature of the thermal-like component C$_\mathrm{Th}$ is obtained by dividing E$_\mathrm{peak}^\mathrm{Th}$ by $\sim$2.5.}
\end{center}
\end{figure*}

\subsection{Evolution of the Thermal-Like and Non-Thermal Components of the C$_\mathrm{nTh}$+C$_\mathrm{Th}$ model}

\subsubsection{The Initial 1 s Plateau}

\begin{figure*}[ht!]
\begin{center}
\includegraphics[totalheight=0.40\textheight, clip]{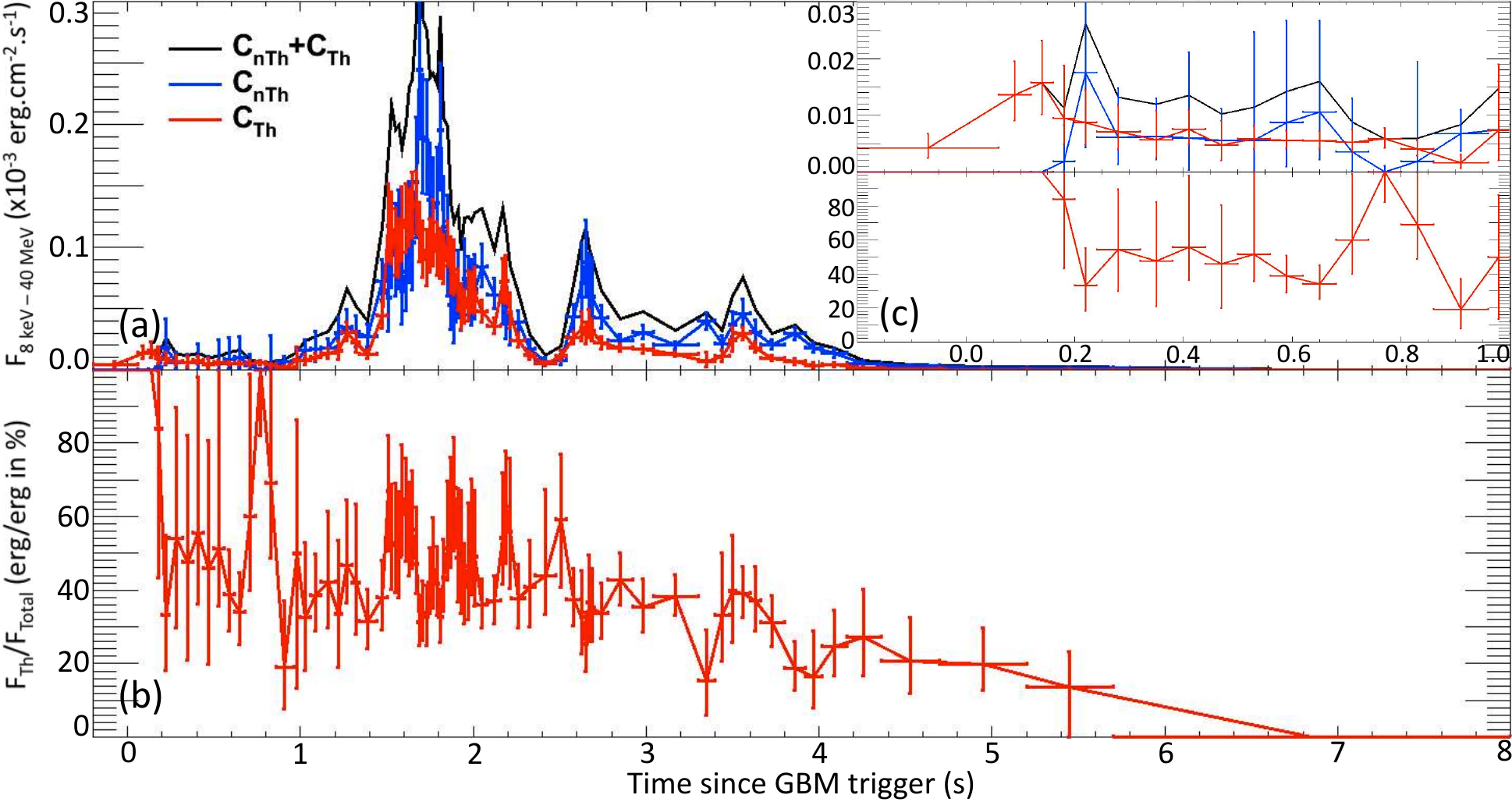}
\caption{\label{fig05}(a) Energy flux evolution between 8 keV and 40 MeV in the context of the C$_\mathrm{nTh}$+C$_\mathrm{Th}$ (with $\alpha_\mathrm{nTh}$=-0.7 and $\alpha_\mathrm{Th}$=+0.6) model. The contributions of the non-thermal (C$_\mathrm{nTh}$) and thermal-like (C$_\mathrm{Th}$) components are displayed in blue and red, respectively. For clarity, no uncertainty is displayed for the total energy flux (black line). (b) Relative contribution of the thermal-like component to the total energy flux between 8 keV and 40 MeV. (c) Zoom of panels a and b during the first second of the emission.}
\end{center}
\end{figure*}

During the initial 1--s plateau (see Figure~\ref{fig01}), the C$_\mathrm{nTh}$+C$_\mathrm{Th}$ model results in a progressive increase of the C$_\mathrm{nTh}$ intensity as shown in Figure~\ref{fig10} but it remains weaker than the intensity of C$_\mathrm{Th}$. The reconstructed photon light curves---in the same energy bands as the count light curves of Figure~\ref{fig01}---resulting from the analysis with the C$_\mathrm{nTh}$+C$_\mathrm{Th}$ model are displayed in Figure~\ref{fig04} and the reconstructed energy light curves between 8 keV and 40 MeV, as well as the contribution of each component to the total energy, are displayed in Figure~\ref{fig05}. In Figure~\ref{fig04}a, the solid red and blue lines and the dashed black line correspond to C$_\mathrm{Th}$, C$_\mathrm{nTh}$ and C$_\mathrm{nTh}$+C$_\mathrm{Th}$, respectively. In Figure~\ref{fig04}b, the reconstructed photon light curves of C$_\mathrm{Th}$ (red) and C$_\mathrm{nTh}$ (blue) are displayed together with C$_\mathrm{Th}$ and C$_\mathrm{nTh}$ $\nu$F$_\nu$ spectral peak energies E$_\mathrm{peak}^\mathrm{Th}$ (orange) and E$_\mathrm{peak}^\mathrm{nTh}$ (cyan), respectively.

While the behavior of E$_\mathrm{peak}^\mathrm{nTh}$ is difficult to identify during the first 1 s of the burst because of large uncertainties, the thermal component C$_\mathrm{Th}$ exhibits a clear monotonic cooling during this period of time (see Figure~\ref{fig02}c and Figure~\ref{fig04}b). E$_\mathrm{peak}^\mathrm{Th}$ decreases from $\sim$800 keV down to $\sim$100 keV, which corresponds to the temperature kT decreasing from $\sim$300 keV down to $\sim$40 keV.

\subsubsection{The Main Emission Episode}

The main emission episode of the burst starts $\sim$1 s after the GBM trigger time. During this phase, the $\gamma$-ray emission strongly increases and exhibits multiple intensity peak structures in the light curves, the most intense one between $\sim$T$_\mathrm{0}$+1.4 and $\sim$T$_\mathrm{0}$+2.4 s (see Figure~\ref{fig01}). As shown in Figures~\ref{fig04} and~\ref{fig05}, the light curves of C$_\mathrm{Th}$ (red) and C$_\mathrm{nTh}$ (blue) undertake globally the same evolution; however, they are not exactly tracking each other, and the intensity of the thermal component peaks before the intensity of the non-thermal one. 
Similarly, the temperature of the thermal component (i.e., $\propto$E$_\mathrm{peak}^\mathrm{Th}$) peaks just before E$_\mathrm{peak}^\mathrm{nTh}$ of the non-thermal one. We note that some intensity peaks in the non-thermal light curves, C$_\mathrm{nTh}$, and in the E$_\mathrm{peak}^\mathrm{nTh}$ variation---such as the one observed between $\sim$T$_\mathrm{0}$+1.9 s and $\sim$T$_\mathrm{0}$+2.2 s in the 600 keV--2 MeV energy range in Figure~\ref{fig04}--b4---are not accompanied by significant increases of the thermal-component temperature and that the variations of E$_\mathrm{peak}^\mathrm{nTh}$ and E$_\mathrm{peak}^\mathrm{Th}$ are not proportional to each other.

After $\sim$T$_\mathrm{0}$+1 s, the temperature, kT,  of the thermal component increases simultaneously with its intensity to reach $\sim$200 keV (i.e., E$_\mathrm{peak}^\mathrm{Th}$$\sim$500 keV) at $\sim$T$_\mathrm{0}+$1.5 s (see Figure~\ref{fig02}c and Figure~\ref{fig04}b). Then the temperature, kT, decreases quite monotonically down to $\sim$10 keV (i.e., E$_\mathrm{peak}^\mathrm{Th}$$\sim$20 keV) with very limited increases matching globally with the non-thermal light curve intensity peaks. Despite the high intensity of the thermal spike at $\sim$T$_\mathrm{0}$+1.5 s---that clearly overpowers the non-thermal emission between 100 and 600 keV (see Figure~\ref{fig04}--a3 \& --b3)---its temperature remains much lower than during the initial plateau that was, however, much fainter.

From T$_\mathrm{0}$+1.44 s to T$_\mathrm{0}$+4.7 s, E$_\mathrm{peak}^\mathrm{nTh}$ clearly tracks the flux of non-thermal component C$_\mathrm{nTh}$; this is particularly evident in the 600 keV to 2 MeV reconstructed photon light curve displayed in Figure~\ref{fig04}--b4. The F$_\mathrm{i}^\mathrm{nTh}$--E$_\mathrm{peak,i}^\mathrm{nTh}$ and L$_\mathrm{i}^\mathrm{nTh}$--E$_\mathrm{peak,i}^\mathrm{nTh,rest}$ relations will be discussed in detail in Section~\ref{sec:relations}.

Overall, the contribution of the thermal component to the total emission is higher at early times, and 
the non-thermal one clearly dominates 
after $\sim$T$_\mathrm{0}$+2 s (see Figure~\ref{fig05}). Because of its narrow shape, C$_\mathrm{Th}$ is particularly intense between 20 and 600 keV (see Figure~\ref{fig04}). The non-thermal component is then globally dominant $<$20 keV and $>$600 keV. 
By comparing the observed count light curves of Figure~\ref{fig01}b displayed in various energy ranges with the reconstructed photon light curves of Figure~\ref{fig04}a displayed in the same energy bands, we can appreciate the very good match of the C$_\mathrm{nTh}$+C$_\mathrm{Th}$ analysis with the data.

\begin{figure*}[ht!]
\begin{center}
\includegraphics[totalheight=0.17\textheight, clip]{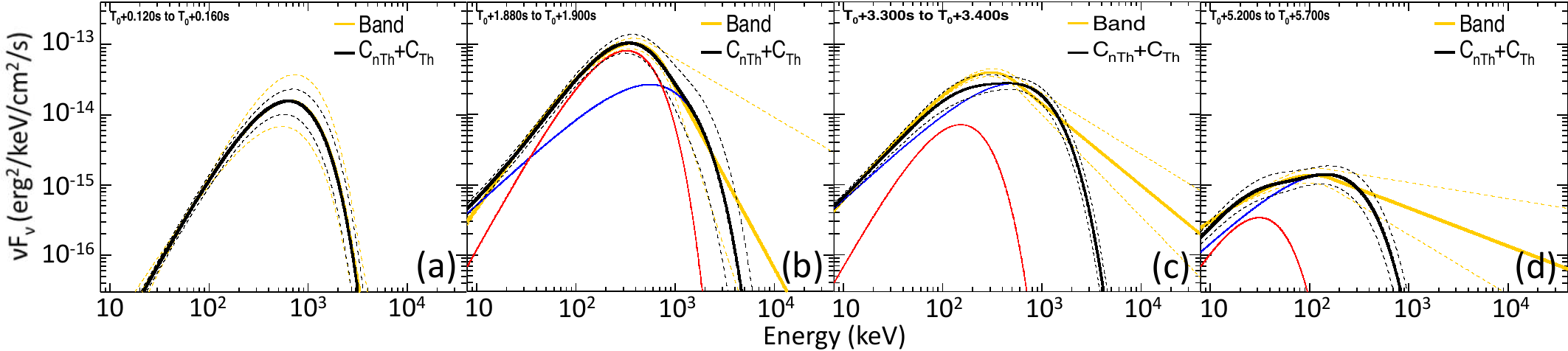}
\caption{\label{fig06}Snapshots of $\nu$F$_\nu$ spectra at various time during the burst---extracted from Figure~\ref{fig10}---exhibiting evolution of the contributions of the non-thermal (C$_\mathrm{nTh}$) and of the thermal-like (C$_\mathrm{Th}$) components of the C$_\mathrm{nTh}$+C$_\mathrm{Th}$ model during the burst. The blue and red lines correspond to C$_\mathrm{nTh}$ and to C$_\mathrm{Th}$, respectively. The solid and dashed black lines correspond to the sum of the two components (i.e., C$_\mathrm{nTh}$+C$_\mathrm{Th}$) and the 1--$\sigma$ uncertainties of the fits using the C$_\mathrm{nTh}$+C$_\mathrm{Th}$ model, respectively. The solid and dashed orange lines correspond to the best fits using a Band function alone and to the 1--$\sigma$ uncertainties of the fits, respectively. The orange lines appear as averages of the black ones. We can particularly note that E$_\mathrm{peak}^\mathrm{Band}$ is displaced towards E$_\mathrm{peak}^\mathrm{Th}$ when C$_\mathrm{Th}$ is intense (panels a \& b) and towards E$_\mathrm{peak}^\mathrm{nTh}$ when C$_\mathrm{Th}$ is fainter (panels c \& d). $\alpha_\mathrm{Band}$ is also biased towards higher values when the thermal-like component is intense. Finally, $\beta_\mathrm{Band}$ is a global average of the high-energy emission and the slope changes have no physical meaning.}
\end{center}
\end{figure*}

\begin{figure*}[ht!]
\begin{center}
\includegraphics[totalheight=0.57\textheight, clip]{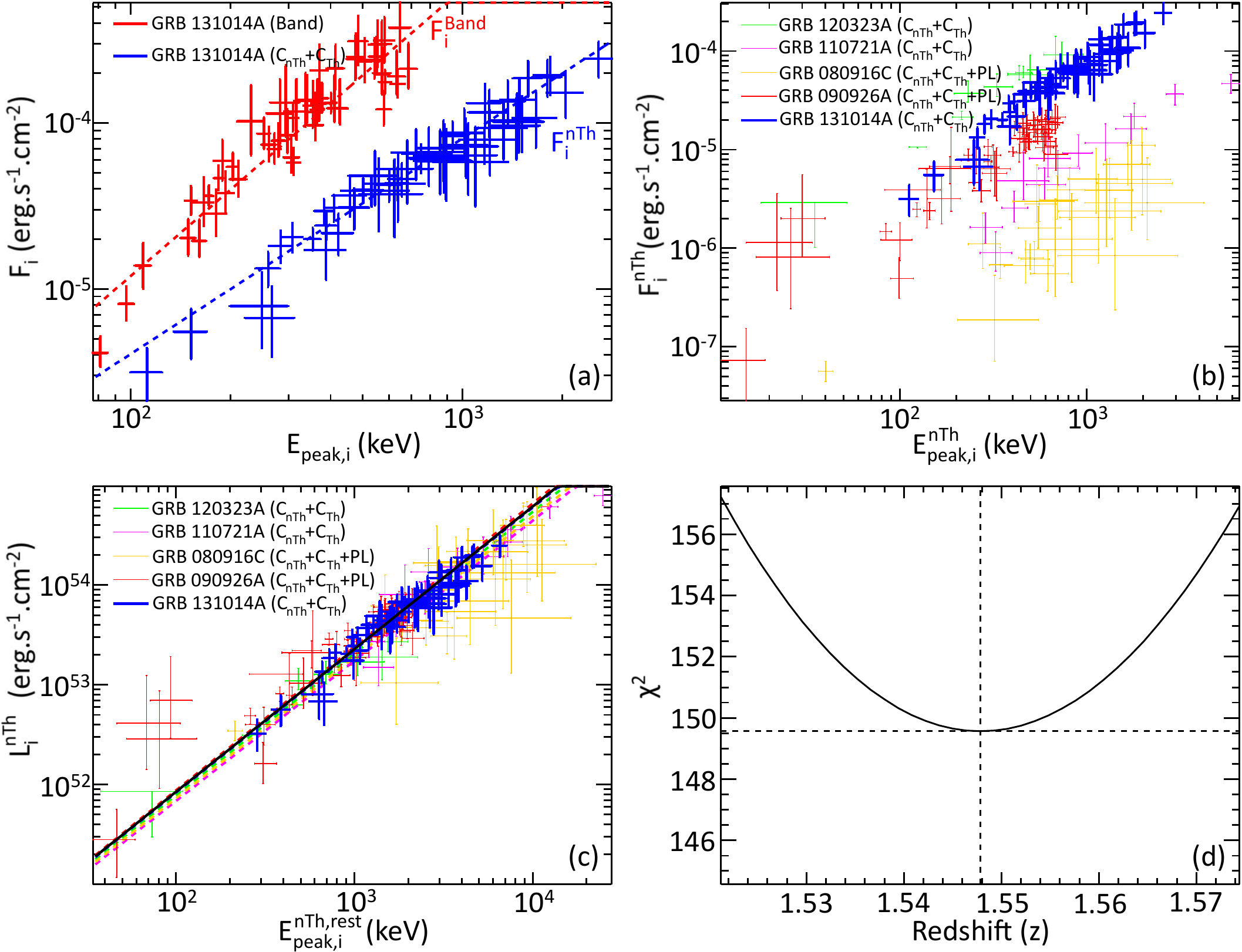}
\caption{\label{fig07}(a) Flux of the Band function,F$_\mathrm{i}^\mathrm{Band}$, versus its $\nu$F$_\nu$ peak energy, E$_\mathrm{peak,i}^\mathrm{Band}$ resulting from the Band-only fits (i.e., F$_\mathrm{i}^\mathrm{Band}$--E$_\mathrm{peak,i}^\mathrm{Band}$ relation) and flux of the non-thermal component, C$_\mathrm{nTh}$, versus its $\nu$F$_\nu$ peak energy, E$_\mathrm{peak,i}^\mathrm{nTh}$ resulting from the C$_\mathrm{nTh}$+C$_\mathrm{Th}$ fits (i.e., F$_\mathrm{i}^\mathrm{nTh}$--E$_\mathrm{peak,i}^\mathrm{nTh}$ relation) in red and blue, respectively. The dashed lines correspond to the PL fit to the data. Only the time intervals after T$_\mathrm{0}$+1.44 s are reported in Figure~\ref{fig07}. (b) F$_\mathrm{i}^\mathrm{nTh}$--E$_\mathrm{peak,i}^\mathrm{nTh}$ relation for GRB~131014A in the context of other GRBs fitted with the new multi-component model presented in~\citet{Guiriec:2013a,Guiriec:2015}. (c) L$_\mathrm{i}^\mathrm{nTh}$--E$_\mathrm{peak,i}^\mathrm{nTh,rest}$ relations for a sample of GRBs with known redshift and for GRB~131014A using the estimated 1.55 redshift value. Each colored dashed line corresponds to the best PL fit to a GRB. The solid dashed black line corresponds to the best PL fit to all GRBs but not GRB~131014A. (d) $\chi^\mathrm{2}$ profile resulting from the minimization in the log--log space---to take into account the uncertainties of the data points on both axis---of the redshift to estimate the value that best matches the time-resolved L$_\mathrm{i}^\mathrm{nTh}$--E$_\mathrm{peak,i}^\mathrm{nTh}$ data points of the GRB~131014A with the L$_\mathrm{i}^\mathrm{nTh}$--E$_\mathrm{peak,i}^\mathrm{nTh}$ relation from Equation~\ref{eq:L-Epeak} (solid black line in panel c) that is derived from a sample of GRBs with known redshift and published in \citet{Guiriec:2013a,Guiriec:2015}. The estimated redshift for GRB~131014A when using the C$_\mathrm{nTh}$+C$_\mathrm{Th}$ model is 1.55$^\mathrm{+0.02}_\mathrm{-0.03}$, which is typical for a long GRB. The same technic applied to the data resulting from the Band-only fits gives a redshift of 0.85$\pm$0.10 (not shown in the Figure.)
}
\end{center}
\end{figure*}

\newpage
\subsection{Comparison of the C$_\mathrm{nTh}$+C$_\mathrm{Th}$ and Band models}
\label{sec:band_C1C2}

As often reported in other GRBs, the fit to the $\gamma$-ray data of GRB~131014A to a Band function alone results in strong variations of the Band spectral parameters during the prompt emission. This is evident in Figure~\ref{fig02} where the evolution of $\alpha_\mathrm{Band}$, $\beta_\mathrm{Band}$ and E$_\mathrm{peak}^\mathrm{Band}$ are reported. $\alpha_\mathrm{Band}$ exhibits strong variation from $\sim$+1.0 down to $\sim$-1 and $\beta_\mathrm{Band}$ values vary from $>$-2.5 down to values making the Band's PL at high energy compatible with an exponential cutoff. It is, however, rather unusual to observe such high values for $\alpha_\mathrm{Band}$, where even the 1--$\sigma$ lower limits can be in many cases $>$0 (see Figure~\ref{fig02}a \& b).

As already reported in \citet{Guiriec:2011,Guiriec:2013a,Guiriec:2015}, the addition of a thermal-like component to the Band function not only improves the fits, but also impacts the parameters of the Band function in a systematic way: in the fits including both the thermal-like and the non-thermal components, both $\alpha_\mathrm{nTh}$ and $\beta_\mathrm{nTh}$ have lower values and E$_\mathrm{peak}^\mathrm{nTh}$ is shifted to much higher energy compared to the fits using a Band function alone (see Table~\ref{tab01}). While the indices of the low-energy PLs resulting from the Band-only fits (i.e., $\alpha_\mathrm{Band}$) are clearly not compatible with non-thermal processes, those resulting from the C$_\mathrm{nTh}$+C$_\mathrm{Th}$ fits (i.e., $\alpha_\mathrm{nTh}$) are quite constant and mostly globally compatible with the -2/3 limit of the synchrotron emission from electrons in the pure slow cooling regime~\citep{Cohen:1997,Crider:1997,Ghisellini:2000}.

The values of $\beta_\mathrm{nTh}$ are usually $<$-3, making them mostly compatible with an exponential cutoff due to the quality of our data.

The discrepancies between the values of E$_\mathrm{peak}^\mathrm{Band}$ (from Band fits) and E$_\mathrm{peak}^\mathrm{nTh}$ (from C$_\mathrm{nTh}$+C$_\mathrm{Th}$ fits) depend on the relative intensities of C$_\mathrm{Th}$ and C$_\mathrm{nTh}$ of C$_\mathrm{nTh}$+C$_\mathrm{Th}$ as illustrated in Figure~\ref{fig06}. When the thermal-like component is very strong, the value of E$_\mathrm{peak}^\mathrm{Band}$ is shifted towards E$_\mathrm{peak}^\mathrm{Th}$; this is for instance the case at early time for GRB~131014A as shown in Figure~\ref{fig06} a \& b. The same result was reported in \citet{Guiriec:2013a} for short GRB~120323A that exhibited a strong thermal component at early time. The weaker the thermal-like component compared to the non-thermal one, the closer E$_\mathrm{peak}^\mathrm{Band}$ is to E$_\mathrm{peak}^\mathrm{nTh}$. This is illustrated by Figure~\ref{fig06} c \& d; indeed, at late times the thermal-like component is subdominant compared to the non-thermal one and E$_\mathrm{peak}^\mathrm{Band}$ is between E$_\mathrm{peak}^\mathrm{Th}$ and E$_\mathrm{peak}^\mathrm{nTh}$. This result is consistent with those reported on long GRBs~100724B, 080916C and 090926A in \citet{Guiriec:2011,Guiriec:2015} as well as at late times in short GRB~120323A in \citet{Guiriec:2013a}. In Figure~\ref{fig02}b we can clearly appreciate the bias of E$_\mathrm{peak}^\mathrm{Band}$ with respect to E$_\mathrm{peak}^\mathrm{nTh}$ and E$_\mathrm{peak}^\mathrm{Th}$. Indeed, at early time---during the first 1 second of the burst---the prompt emission of GRB~131014A is nearly purely thermal and, therefore, E$_\mathrm{peak}^\mathrm{Band}$ is equal to E$_\mathrm{peak}^\mathrm{Th}$; the orange and red lines overlap perfectly showing a monotonic decrease of E$_\mathrm{peak}^\mathrm{Band}$ matching with the cooling of the thermal-like component (i.e., $\propto$E$_\mathrm{peak}^\mathrm{Th}$.) After $\sim$T$_\mathrm{0}$+1 s, the discrepancies between E$_\mathrm{peak}^\mathrm{Band}$ and E$_\mathrm{peak}^\mathrm{Th}$ increase as the relative intensity of the non-thermal component compared to the thermal one gets larger; as a result, E$_\mathrm{peak}^\mathrm{Band}$ (orange line) gets closer to E$_\mathrm{peak}^\mathrm{nTh}$ (red line).

To conclude, if we consider that the intrinsic shape of the $\nu$F$_\nu$ spectrum is adequately described with the two humps of the C$_\mathrm{nTh}$+C$_\mathrm{Th}$ model whose relative intensities and positions (i.e., E$_\mathrm{peak}^\mathrm{Th}$ and E$_\mathrm{peak}^\mathrm{nTh}$) evolve with time, it is natural that the use of a simpler model involving a single hump such as the Band function results in an average of the most complex shape with possible strong variations of the spectral parameters of the simplest model. The use of the simplest model would then hide the intrinsic physics relevant to the prompt emission even if it remains overall a statistically good fit. This is clearly illustrated in Figures \ref{fig06} and~\ref{fig10} where the C$_\mathrm{nTh}$+C$_\mathrm{Th}$ and the Band models are overplotted.

\subsection{F$^{nTh}_{i}$--E$_{peak,i}^{nTh}$ and L$^{nTh}_{i}$--E$^{nTh,rest}_{peak,i}$ relations, and redshift estimate for GRB~131014A}
\label{sec:relations}

In \citet{Guiriec:2013a,Guiriec:2015} we derived a new relation---that we believe to be intrinsic to every GRB---between the time-resolved flux of the non-thermal component, F$_\mathrm{i}^\mathrm{nTh}$, and its $\nu$F$_\nu$ peak energy, E$_\mathrm{peak,i}^\mathrm{nTh}$ (i.e., F$_\mathrm{i}^\mathrm{nTh}$--E$_\mathrm{peak,i}^\mathrm{nTh}$ relation). The F$_\mathrm{i}^\mathrm{nTh}$--E$_\mathrm{peak,i}^\mathrm{nTh}$ relations have about the same index value (i.e., $\sim$+1.4) for all GRBs when fitted to a PL. Moreover, when corrected for the redshift (i.e., in the source frame), a possible universal relation common to all GRBs---including both short and long ones---appears between the luminosity of the non-thermal component, L$_\mathrm{i}^\mathrm{nTh}$, and its $\nu$F$_\nu$ peak energy, E$_\mathrm{peak,i}^\mathrm{nTh,rest}$:

\begin{equation}
\label{eq01}
\mathrm{L_i^{nTh}}=\mathrm{(9.6\pm1.1)\times10^{51}~\left(\frac{E_{peak,i}^{nTh,rest}}{100~keV}\right)^{1.38\pm0.04}~erg.s^{-1}}
\label{eq:L-Epeak}
\end{equation}
from \citet{Guiriec:2015}.

It was noticeable that, for every burst, the very first instants of the prompt emission during the initial intensity rising phase may not follow this relation but rather a possible anti-correlation; however, after this initial rising phase, the relation was valid and the same for all the other rising and decaying emission episodes.

Figure~\ref{fig07}a shows the F$^\mathrm{Band}_\mathrm{i}$--E$_\mathrm{peak,i}^\mathrm{Band}$ and F$^\mathrm{nTh}_\mathrm{i}$--E$_\mathrm{peak,i}^\mathrm{nTh}$ relations obtained when fitting Band-alone and C$_\mathrm{nTh}$+C$_\mathrm{Th}$, respectively, to the time-resolved data of GRB~131014A after T$_\mathrm{0}$+1.44 s. The PL indices of the F$^\mathrm{Band}_\mathrm{i}$--E$_\mathrm{peak,i}^\mathrm{Band}$ and F$^\mathrm{nTh}_\mathrm{i}$--E$_\mathrm{peak,i}^\mathrm{nTh}$ relations for GRB~131014A are +1.92$\pm$0.07 and +1.43$\pm$0.03, respectively; interestingly, the index resulting from the C$_\mathrm{nTh}$+C$_\mathrm{Th}$ fits is perfectly consistent with those reported in \citet{Guiriec:2013a,Guiriec:2015} for the analysis of GRBs 120323A, 110721A, 080916C and 090926A, but not the index obtained with Band-only. We also observe that  E$_\mathrm{peak}$ reaches much higher values in the case of C$_\mathrm{nTh}$+C$_\mathrm{Th}$ than in the case of Band-alone as already mentioned in Section~\ref{sec:band_C1C2}, and that the flux of the non-thermal component is much lower in the case of C$_\mathrm{nTh}$+C$_\mathrm{Th}$ than in the case of Band-alone because of the very intense thermal-like component. In most GRBs, the thermal-like component only accounts for a few percent of the total emission and there are, therefore, no strong discrepancies in term of flux between Band-alone and C$_\mathrm{nTh}$ from C$_\mathrm{nTh}$+C$_\mathrm{Th}$; the discrepancies usually concern the E$_\mathrm{peak}$ values that can be very different in the Band-only and the C$_\mathrm{nTh}$+C$_\mathrm{Th}$ models. We can also notice the scatter of the data points in the Band-only scenario that is much stronger than in the case of the C$_\mathrm{nTh}$+C$_\mathrm{Th}$ one.

Unfortunately there is no redshift measurement for GRB~131014A nor any host galaxy identification, and it is, therefore, impossible to verify if GRB~131014A is compatible with the possible universal L$^\mathrm{nTh}_\mathrm{i}$--E$^\mathrm{nTh,rest}_\mathrm{peak,i}$ relation presented in Equation~\ref{eq:L-Epeak} and derived from a sample of bursts with known redshift \citep{Guiriec:2013a,Guiriec:2015}.

If we assume that the L$^\mathrm{nTh}_\mathrm{i}$--E$^\mathrm{nTh,rest}_\mathrm{peak,i}$ relation, which is intrinsic to the non-thermal component, is indeed universal, then we must also accept that C$_\mathrm{nTh}$+C$_\mathrm{Th}$ is the closest model to the ``true'' physics of GRB~131014A since only the L$^\mathrm{nTh}_\mathrm{131014A,i}$--E$^\mathrm{nTh,rest}_\mathrm{peak,i}$ relation devised from the C$_\mathrm{nTh}$+C$_\mathrm{Th}$ model has a compatible slope with Equation~\ref{eq:L-Epeak} and not the L$^\mathrm{Band}_\mathrm{131014A,i}$--E$^\mathrm{rest,Band}_\mathrm{peak,i}$ one. Figure~\ref{fig07}b puts GRB~131014A in the context of other GRBs with measured redshifts.  The data of GRB~131014A (blue) lie in between those of GRB~120323A (green) and those of GRB~090926A (red). GRB~120323A is a short GRB without measured redshift but the Swift short GRB sample has a typical redshift of $\sim$1 and GRB~090926A has a spectroscopic redshift measurement of $\sim$2.1; therefore, a quick estimate indicates that the redshift of GRB~131014A is in between 1 and 2.1.

We performed a more quantitative estimate of the redshift of GRB~131014A; to do so, we minimized the distance between the L$_\mathrm{i}^\mathrm{nTh}$--E$_\mathrm{peak,i}^\mathrm{nTh,rest}$ data of GRB~131014A--resulting from the C$_\mathrm{nTh}$+C$_\mathrm{Th}$ analysis--and Equation~\ref{eq:L-Epeak} in varying the redshift and in using a linear fit in the log--log space to take into account the uncertainties on both L$_\mathrm{i}^\mathrm{nTh}$ and E$_\mathrm{peak,i}^\mathrm{nTh,rest}$ quantities. The $\chi^\mathrm{2}$ profile resulting from the fit is plotted in Figure~\ref{fig07}d and shows that a redshift of $\sim$1.55
---within 1--$\sigma$ uncertainties---is required for GRB~131014A to be compatible with the L$^\mathrm{nTh}_\mathrm{i}$--E$^\mathrm{nTh,rest}_\mathrm{peak,i}$ relation from Equation~\ref{eq:L-Epeak}. This redshift estimate is quite typical of long GRBs; indeed, the redshift distribution for long GRBs peaks at $\sim$2.16~\citep{Jakobsson:2012}. We must note that this quick estimate does not take into account the dispersion of Equation~\ref{eq:L-Epeak} but it only use the central values for the parameters; however, based on the uncertainties on the parameters of the L$^\mathrm{nTh}_\mathrm{i}$--E$^\mathrm{nTh,rest}_\mathrm{peak,i}$ relation, we do not expect strong discrepancies. A specific follow-up article will be dedicated to the study of the F$_\mathrm{i}^\mathrm{nTh}$--E$_\mathrm{peak,i}^\mathrm{nTh}$ and L$_\mathrm{i}^\mathrm{nTh}$--E$_\mathrm{peak,i}^\mathrm{nTh,rest}$ relations as well as to their use as a GRB redshift estimator and as tools for cosmology. Figure~\ref{fig07}c shows the L$^\mathrm{nTh}_\mathrm{i}$--E$^\mathrm{nTh,rest}_\mathrm{peak,i}$ relations for a sample of bursts including GRB~131014A for which we used the redshift of 1.55 that we just determined.  We note the very good overlap of the data. The color dashed lines correspond to the best PL fits to each burst individually, and the solid black line corresponds to the best PL fit to the whole sample together but excluding GRB~131014A.

The same technique applied to the Band-only fits results in a redshift of $\sim$0.85
---instead of $\sim$1.55 
obtained with C$_\mathrm{nTh}$+C$_\mathrm{Th}$---that is in the low-value tail of the long-GRB redshift distribution, and therefore less typical.


\section{Interpretation}
\label{section:interpretation}

In the following, we discuss our observational results for C$_\mathrm{nTh}$ and C$_\mathrm{Th}$ within the framework of the popular  ``standard" fireball  photosphere scenario with non-thermal emission.
In this scenario C$_\mathrm{Th}$ represents the quasi-thermal photospheric component and the non-thermal component, C$_\mathrm{nTh}$, arises from dissipation taking place above the photosphere, e.g. from the synchrotron emission produced by non-thermal dissipative processes (e.g., internal shocks or magnetic reconnections) above the photosphere~\citep[e.g.,][]{Sari:1997,Daigne:1998,Piran:1999,Hascoet:2013}. Other possible interpretations, such as the dissipative photospheric models---where C$_\mathrm{Th}$ would result from a moderate or intermittent subphotospheric dissipation and C$_\mathrm{nTh}$ would correspond to a reprocessed thermal emission below the photosphere via inverse Compton mechanisms~\citep[e.g.,][]{Peer:2006,Beloborodov:2010,Lazzati:2013,Vurm:2015}---are not discussed in this article. Our goal is to evaluate the properties of the relativistic outflow of GRB 131014A in the context of the ``standard'' fireball model but we do not discard any other possible interpretations and we encourage the comparison of our observational model with the predictions of other theoretical scenarios.


\subsection{The Photospheric Emission}

\begin{figure}[ht!]
\begin{center}
\includegraphics[totalheight=0.30\textheight, clip]{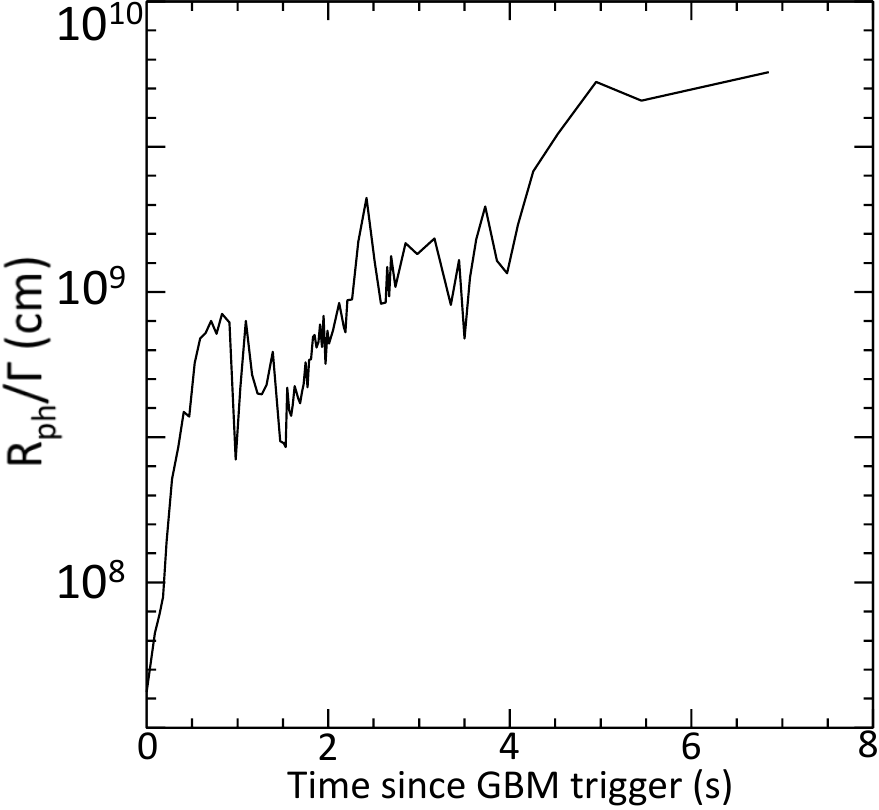}
\caption{\label{fig08}Evolution of the ratio R$_\mathrm{ph}$/$\Gamma$ as given by Equation~\ref{eq01b} for z=1.55 as proposed in Section~\ref{sec:relations}.}
\end{center}
\end{figure}
\vspace{-0.6cm}

C$_\mathrm{Th}$ can be viewed as the footprint of the jet photosphere. The relative evolution 
of its energy flux, F$_\mathrm{Th}$, and the observed temperature, T$_\mathrm{obs}$, then allows to obtain an estimate of the apparent photospheric radius 
\begin{equation}
\label{eq01b}
{R_\mathrm{ph}\over \Gamma}\sim D_{\rm L}^{1/2}\,\left({F_\mathrm{Th}\over
\sigma}\right)^{1/2}{1\over [(1+z)T_{\rm obs}]^2}
\end{equation}
---with $\sigma$ the Stefan-Boltzmann constant---which is displayed in Figure~\ref{fig08} for z=1.55 as proposed in Section~\ref{sec:relations}. Unfortunately we can only obtain in a model independent way the ratio $R_\mathrm{ph}/\Gamma$. Further interpretation of this result depends on the question whether there is dissipation or not below the photosphere.

\subsection{The ``standard" fireball scenario and non-thermal dissipation above the photosphere}

\begin{figure*}
\begin{center}
\includegraphics[totalheight=0.63\textheight, clip]{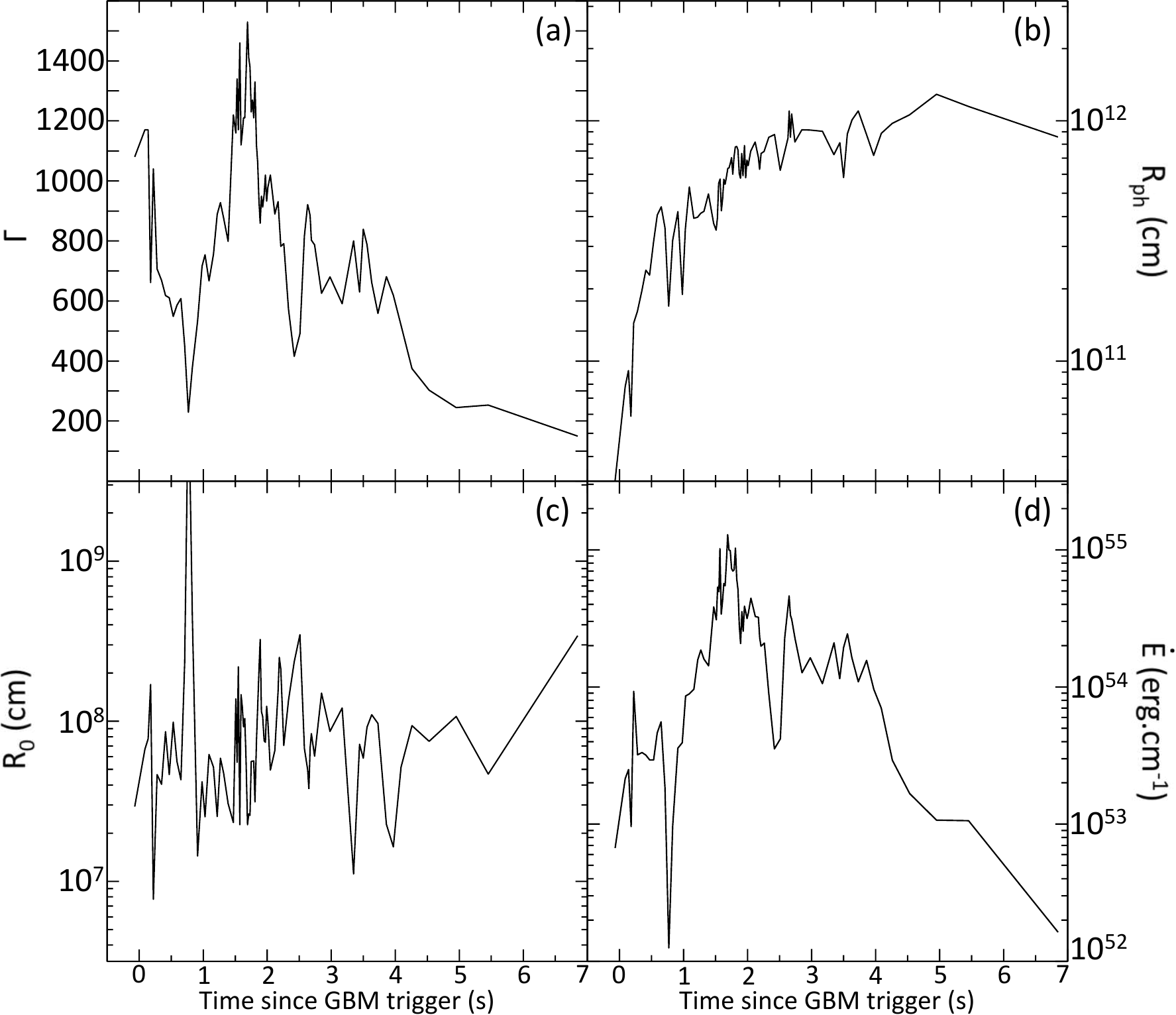}
 \end{center}
\caption {\label{fig09}Lorentz factor, $\Gamma$, photospheric radius, R$_\mathrm{ph}$, injection radius, R$_\mathrm{0}$, and injected 
power, ${\dot E}$, as a function of time for GRB 131014A for z=1.55.}
\end{figure*}

Within this scenario
the thermal and non-thermal components C$_\mathrm{Th}$ and C$_\mathrm{nTh}$ are truly separate contributions.
In this case the standard fireball model~\citep{Piran:1999} provides a natural framework for the interpretation of the 
thermal component, C$_\mathrm{Th}$, as the residual thermal emission of the thermal fireball. Assuming that no dissipation takes place below the photosphere and that the jet propagation is conical we can 
go beyond Equation~\ref{eq01b} and separately provides
the photopheric radius, $R_{\rm ph}$, 
the bulk Lorentz factor, $\Gamma$, as well as the radius at the basis of the flow, $R_0$,
and the injected power, ${\dot E}$, as a function of observer time.
They are given by~\citep[see Equations 3 to 7 in][]{Guiriec:2013a}
\begin{equation}
\label{eq02}
R_{\rm ph}=\left[{\sigma\over 16 m_p c^2}{D_L^5 F_{Th} {\cal R}^3
\over (1+z)^6}{1-R\over R}\right]^{1/4}\times [(1+\sigma_{\rm M})f_{\rm nTh}]^{-1/4}
\end{equation} 
\begin{equation}
\label{eq03}
\Gamma=\left[{\sigma\over m_p c^2}{(1+z)^2 D_L F_{Th} 
\over {\cal R}}{1-R\over R}\right]^{1/4}\times [(1+\sigma_{\rm M})f_{\rm nTh}]^{-1/4}
\end{equation} 
\begin{equation}
\label{eq04}
R_0=\left[{D_L {\cal R}\over 2 (1+z)^2}\left({R\over 1-R}\right)^{3/2}
\right]\times \left[{f_{\rm nTh}\over \epsilon_{\rm Th}}\right]^{3/2} 
\end{equation} 
\begin{equation}
\label{eq05}
{\dot E}=\left[4\pi D_{\rm L}^2 F_{\rm Th} {1-R\over R}\right]\times f_{\rm nTh}^{-1}
\end{equation} 
where m$_\mathrm{p}$ is the proton mass, $F_{\rm Th}$ is thermal energy flux, $R=F_{\rm Th}/F_{\rm Tot}$ (with F$_{\rm Tot}$ the total energy flux in both C$_\mathrm{nTh}$ and C$_\mathrm{Th}$) and $\cal R$ 
is given by
\begin{equation}
\label{eq06}
{\cal R}=\left({F_{\rm Th}\over \sigma T_{obs}^4}\right)^{1/2}\ .
\end{equation} 
In Equations~\ref{eq02} to~\ref{eq05} the factors on the left are directly
obtained from the data while some assumptions have to be made 
for those on the right.  
The quantity $[(1+\sigma_\mathrm{M})f_{\rm nTh}]^{-1/4}$ can be expected to be not too
different from unity ($\sigma_\mathrm{M}$ is the magnetization, $f_{\rm nTh}$ the efficiency of the non thermal process).  The situation is less
clear for Equations~\ref{eq04} and~\ref{eq05}. We have adopted 
${f_{\rm nTh}\over \epsilon_{\rm Th}}=0.3$ ($\epsilon_{\rm Th}$ 
being the fraction of the initial energy released by the source in thermal form).

The evolution of $\Gamma$, $R_{\rm ph}$, $R_0$ and ${\dot E}$ 
resulting from the data is then
shown in Figure~\ref{fig09} (assuming that the burst is located at $z=1.55$ as proposed in Section~\ref{sec:relations}).  
It can be noticed that: 
\begin{itemize}
\item
The photospheric radius increases steadily. After an initial sharp
rise during the first second, it increases more slowly (by a factor of about 3)
until the end of the prompt phase at $\sim$T$_\mathrm{0}$+7 s. Since $R_{\rm ph}\sim
{\dot E}/\Gamma^3$, this suggests that the Lorentz factor approximately behaves
as ${\dot E}^{1/3}$ after the first second. 
One notices that the values obtained for the Lorentz factor and the 
injected power are quite large. 
More ``comfortable'' values would be obtained 
for a closer burst, with $z$ being for example divided by 2 (a factor 1.6 smaller for the Lorentz factor, nearly a factor of 10 for ${\dot E}$).   
\item
The injection radius $R_0$ stays confined within an order of magnitude,
which is satisfactory since this is not a priori imposed in the calculation. 
However the absolute value appears somewhat large (but this might be acceptable if the jet is not conical while it propagates within the stellar envelope). 
We note that our estimates of $\Gamma$ and R$_\mathrm{0}$ are in agreement with those reported in~\citet{Peer:2007} using time-integrated spectra of BATSE GRBs.
\end{itemize}

In this model the non-thermal component, 
C$_\mathrm{nTh}$,  arises from non-thermal dissipation above the photosphere. A possible interpretation is synchrotron emission produced by charged particles accelerated above the photosphere via internal 
shocks for instance or magnetic reconnection. Interestingly once the thermal component is taken into account the low energy slope of C$_\mathrm{nTh}$ does not contradict the ``synchrotron line of death" \citep[e.g.,][]{Cohen:1997,Crider:1997,Ghisellini:2000} as would have been the case if the spectrum was interpreted in terms of a single Band function. The correlation between C$_\mathrm{Th}$
and C$_\mathrm{nTh}$ is reasonable within this model as one expect the residual thermal photospheric component to follow the overall energy of the outflow and hence  C$_\mathrm{nTh}$. With the large Lorentz factor, the time delay between emission of the non-thermal component at say $10^{14}$ cm and the emission at the photosphere at $10^{12}$cm is insignificant.

\section{Summary}
\label{section:conclusion}

GRB 131014A is a rare event for its prompt emission: not only it is a bright GRB but it also exhibits an unusually intense thermal-like component. GRB 131014A is an ideal laboratory for studying in detail the prompt emission thermal-like component, which has now been detected in numerous GRBs, and its connection to the non-thermal one:
\begin{itemize}
\item Long GRB 131014A exhibits an initial short pure thermal emission episode. After $\sim$160 ms, a non-thermal component is detected and its intensity progressively increases until overpowering the thermal one at late time. While the thermal component usually contributes less than $\sim$20\% of the total energy, here it reaches $\sim$50\%.
\item The thermal component monotonically cools from kT$\sim$300 keV down to kT$\sim$40 keV during an initial $\sim$1 s low-intensity light-curve plateau. Then the temperature of the thermal component rises together with its flux; however, despite a much higher flux at $\sim$T$_\mathrm{0}$+1.5 s than during the initial plateau, the temperature only reaches kT$\sim$200 keV. After $\sim$T$_\mathrm{0}$+1.5 s, the thermal component globally cool monotonically down to kT$\sim$9 keV.
\item  Despite some similarities in their global variation, neither the flux nor the $\nu$F$_\nu$ spectral peak energy of the thermal-like and of the non-thermal components track each other perfectly.
\item The strong variations of the spectral parameters when fitting Band alone to the data may be due to the presence of simultaneous multiple components, the resulting Band function being only the average of the total emission without strong physical meaning.
\item The new model proposed in~\citet{Guiriec:2015} including a thermal-like and a non-thermal component (i.e., C$_\mathrm{nTh}$+C$_\mathrm{Th}$) fits the data of GRB~131014A at least as well as a Band function alone and with the same number of free parameters. The low-energy power-law spectral index of C$_\mathrm{nTh}$ remains constant around -0.7 during the whole burst duration. The high-energy power-law spectral index of C$_\mathrm{nTh}$ is $<$3.5 and compatible with an exponential cutoff according to the quality of our data set. The spectral shape of the thermal-like component, C$_\mathrm{Th}$, is broader than a pure black-body and has a spectral index of about +0.6 instead of the +1 of the Planck function.
\item When fitting C$_\mathrm{nTh}$+C$_\mathrm{Th}$ to the data, a strong correlation appears between the time-resolved energy flux of the non-thermal component, F$_\mathrm{i}^\mathrm{nTh}$, and its $\nu$F$_\nu$ peak energy, E$_\mathrm{peak,i}^\mathrm{nTh}$ (i.e., F$_\mathrm{i}^\mathrm{nTh}$--E$_\mathrm{peak,i}^\mathrm{nTh}$ relation). The power-law index of the F$_\mathrm{i}^\mathrm{nTh}$--E$_\mathrm{peak,i}^\mathrm{nTh}$ relation of GRB 131014A is identical to the one reported for other GRBs in \citet{Guiriec:2013a,Guiriec:2015}. If the L$_\mathrm{i}^\mathrm{nTh}$--E$_\mathrm{peak,i}^\mathrm{nTh,rest}$ relation is universal as proposed in \citet{Guiriec:2013a,Guiriec:2015}, GRB 131014A would have a redshift of $\sim$1.55 that is a typical value for long GRBs.
\end{itemize}

The observational results of GRB 131014A can be nicely interpreted with models in which the non-thermal emission originates at radii much larger than the photosphere radius.
In the context of the ``standard'' fireball model, the evolution of the photospheric radius in GRB 131014 appears reasonable
and the radius at the base of the flow remains constant as expected. However, the radius at the base of the flow as well as the injected power have slightly high values and a redshift of $\sim$0.8---instead of our estimate of $\sim$1.55---would provide more ``comfortable'' values. These results do not discard the possibility that other theoretical interpretations may explain our proposed observational model for GRB 131014A as well, and we encourage further comparison.

\section{Acknowledgments}
To complete this project, S.G. was supported by the NASA Postdoctoral Program (NPP) at the NASA/Goddard Space Flight Center, administered by Oak Ridge Associated Universities through a contract with NASA as well as by the NASA grants NNH11ZDA001N and NNH13ZDA001N awarded to S.G. during the cycles 5 and 7 of the NASA Fermi Guest Investigator Program. We thank Andrei Beloborodov for his very useful comments.


\newpage
\begin{appendix}
\label{appendix}

 Table~\ref{tab01} contained the values of the spectral parameters together with their 1--$\sigma$ uncertainties resulting from the fine-time analysis fits to the various models tested in this article.

Figure~\ref{fig10} reports the results of the fine-time spectral analysis of GRB~131014A using either a Band function alone or the combination of a thermal-like and a non-thermal component.

\end{appendix}

\newpage
\setcounter{figure}{0}
\renewcommand{\thefigure}{A\arabic{figure}}

\begin{figure*}
\begin{center}

\includegraphics[totalheight=0.98\textheight, clip]{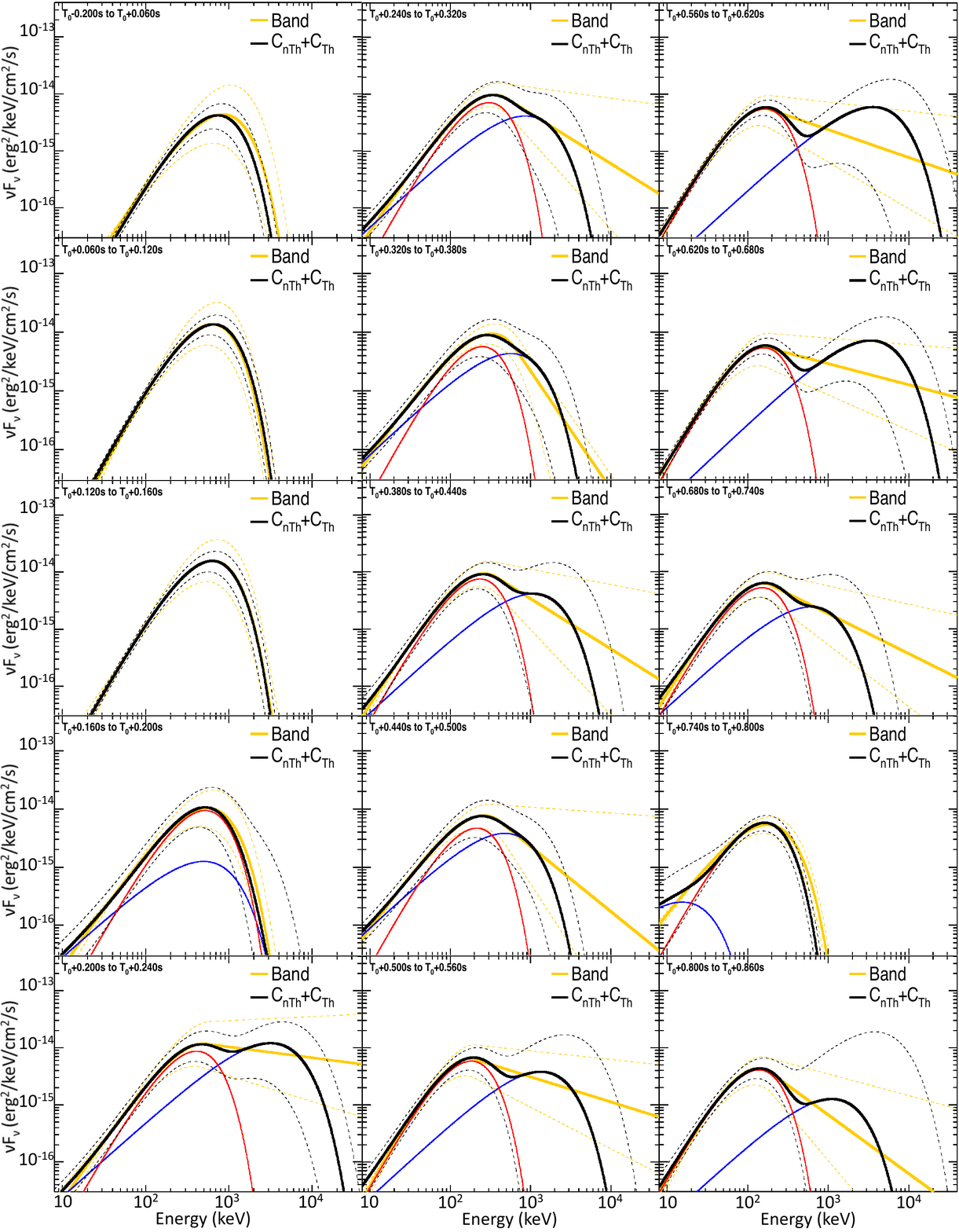}

\end{center}
\end{figure*}

\newpage

\begin{figure*}
\begin{center}

\includegraphics[totalheight=0.98\textheight, clip]{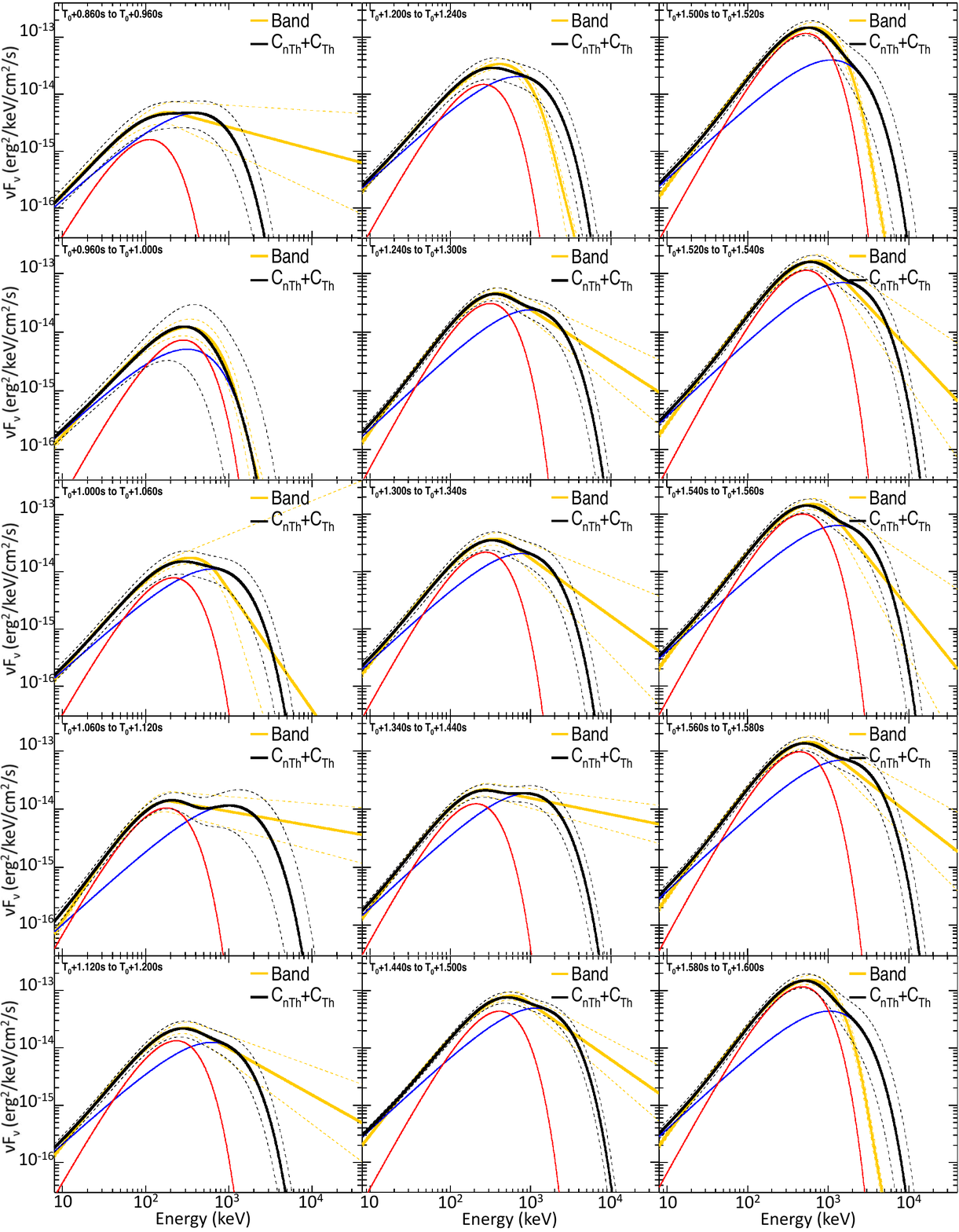}

\end{center}
\end{figure*}

\newpage

\begin{figure*}
\begin{center}

\includegraphics[totalheight=0.98\textheight, clip]{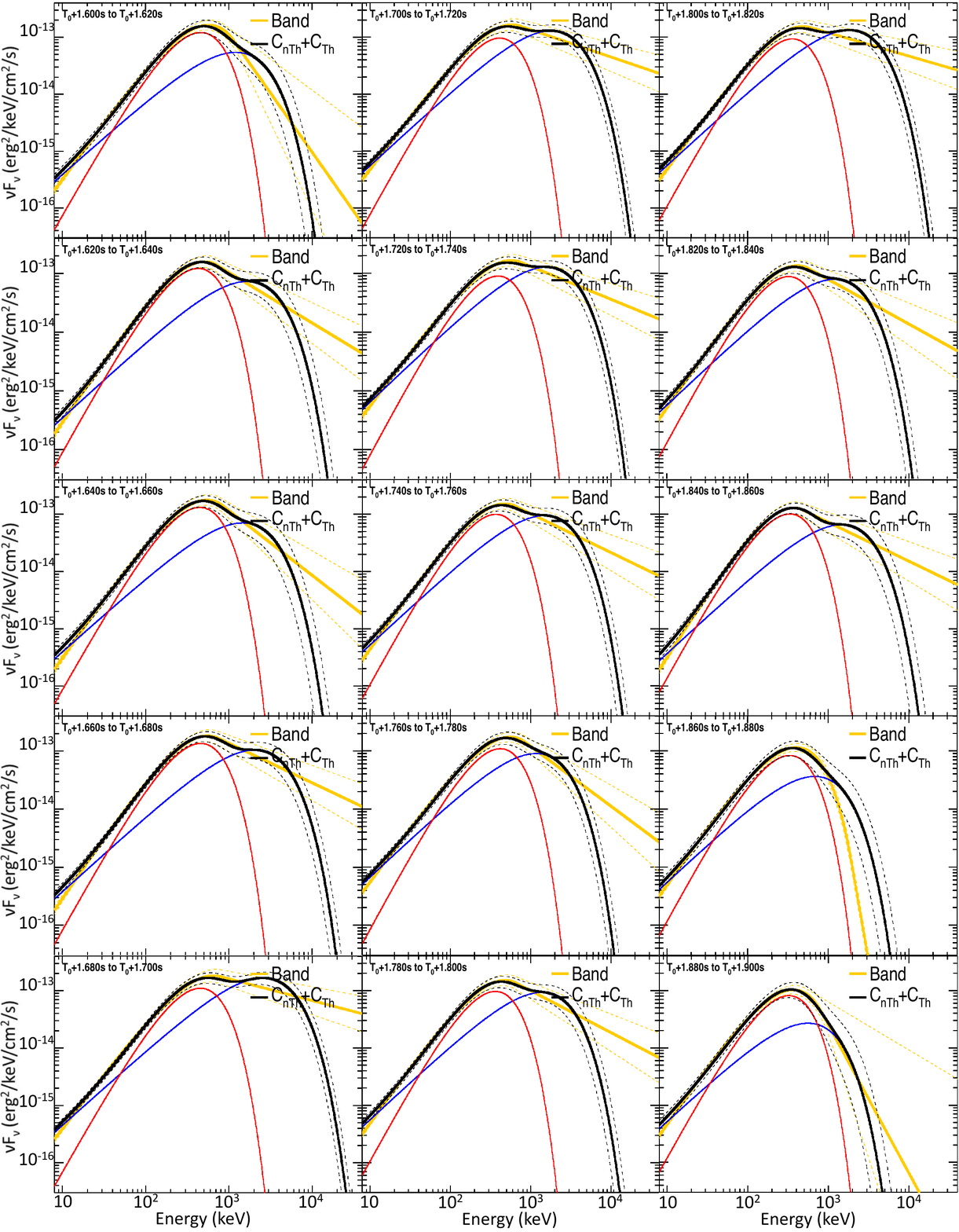}

\end{center}
\end{figure*}

\newpage

\begin{figure*}
\begin{center}

\includegraphics[totalheight=0.98\textheight, clip]{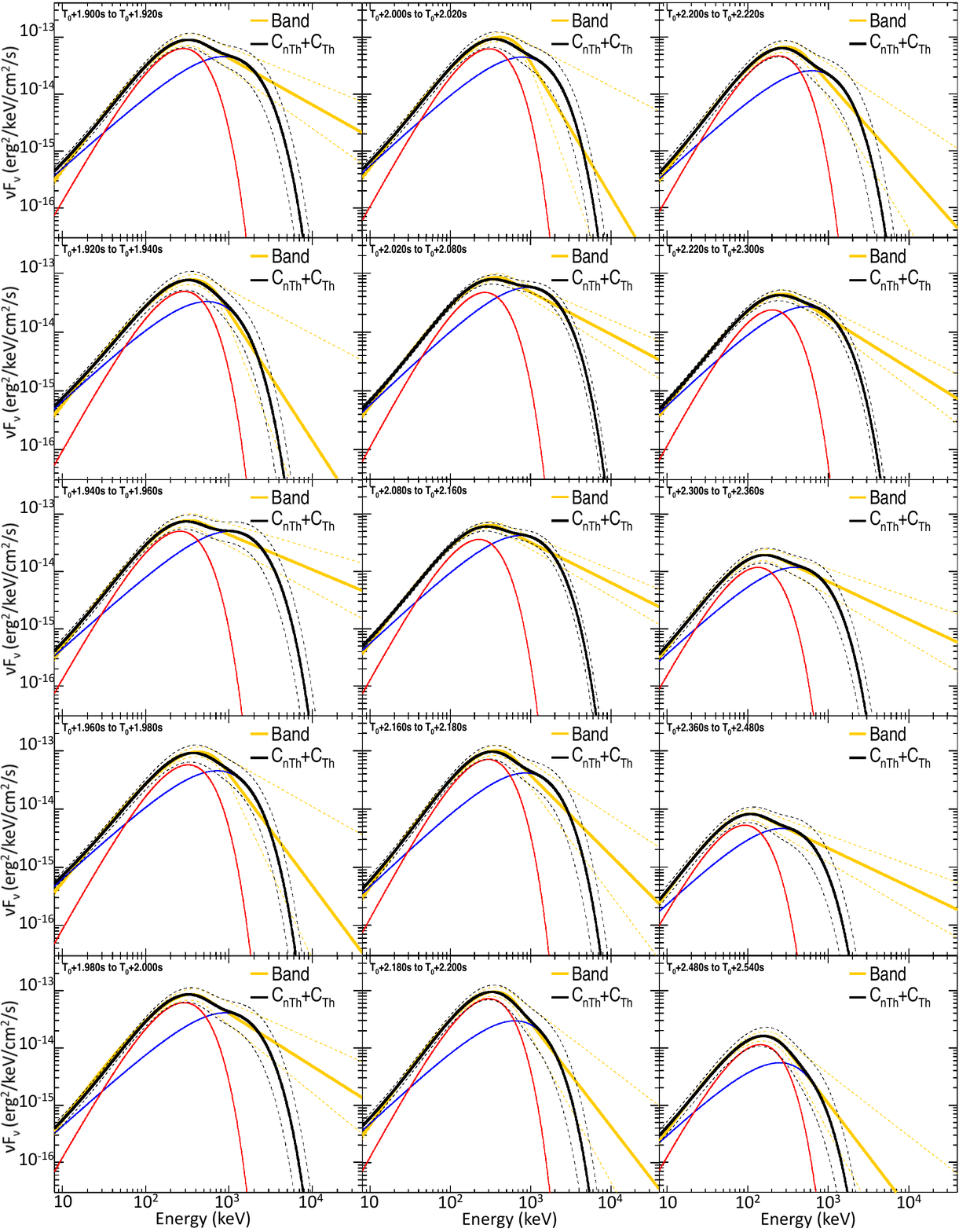}

\end{center}
\end{figure*}

\newpage

\begin{figure*}
\begin{center}

\includegraphics[totalheight=0.98\textheight, clip]{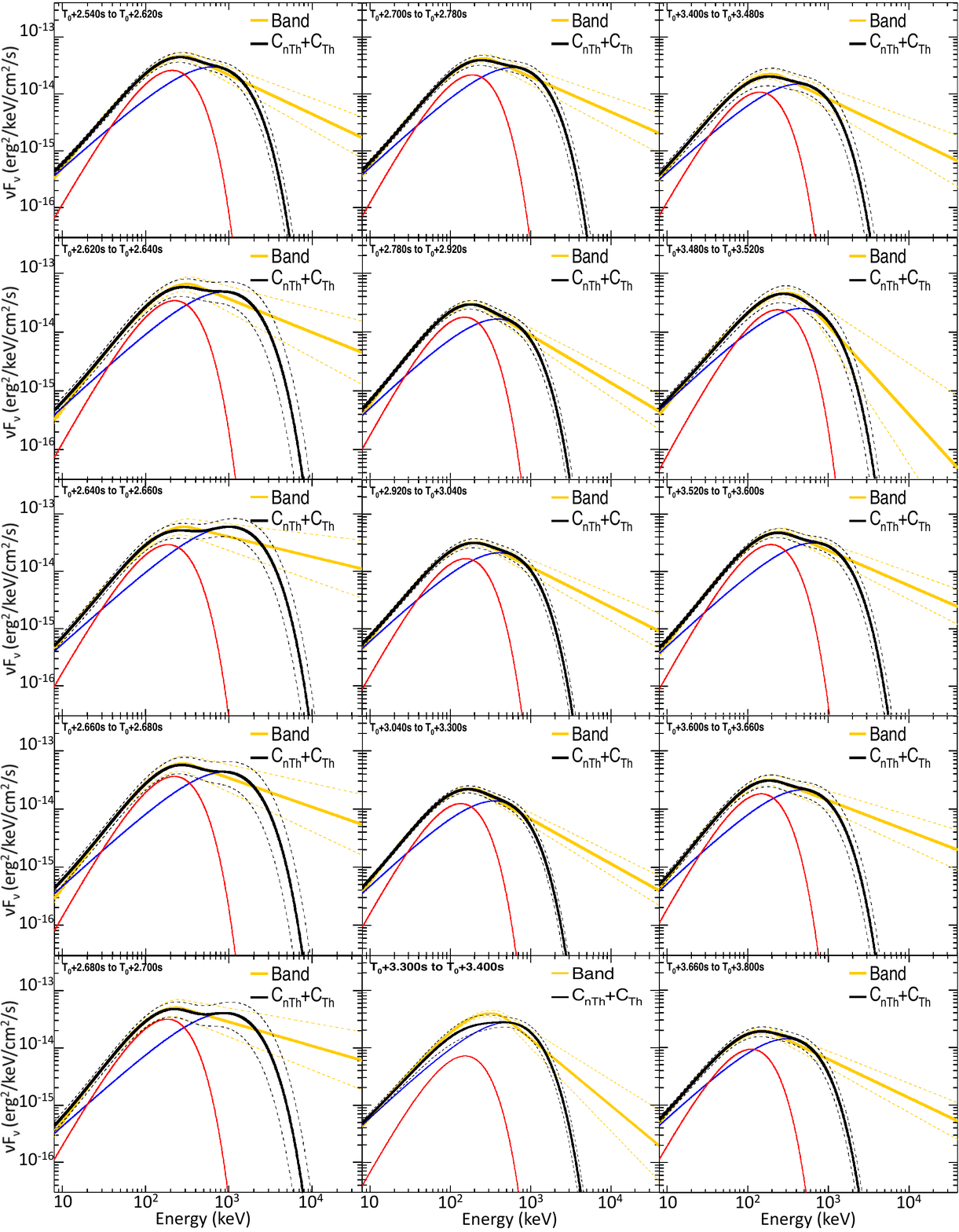}

\end{center}
\end{figure*}

\newpage

\begin{figure*}
\begin{center}

\includegraphics[totalheight=0.6\textheight, clip]{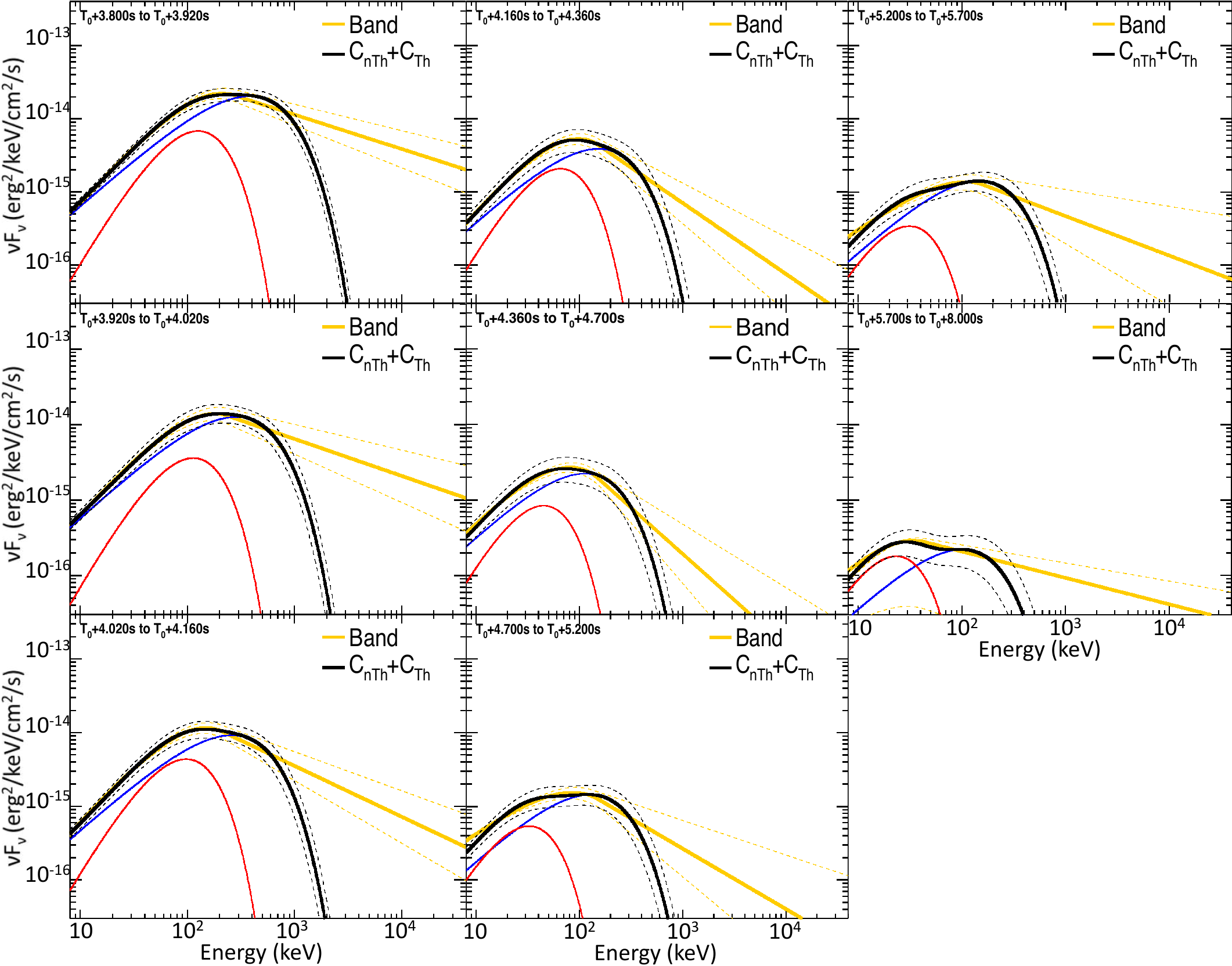}

\caption{\label{fig10}$\nu$F$_\nu$ spectra fits from the fine-time analysis. The solid yellow and black lines correspond to the best Band-only and C$_\mathrm{nTh}$+C$_\mathrm{Th}$ (with $\alpha_\mathrm{nTh}$=-0.7 and $\alpha_\mathrm{Th}$=+0.6) fits, respectively. The dashed yellow and black lines correspond to the 1--$\sigma$ confidence region of the Band-only and C$_\mathrm{nTh}$+C$_\mathrm{Th}$ fits, respectively. The solid blue and red lines correspond to the C$_\mathrm{nTh}$ and C$_\mathrm{Th}$ components of C$_\mathrm{nTh}$+C$_\mathrm{Th}$, respectively.}
\end{center}
\end{figure*}

\newpage
\setcounter{table}{0}
\renewcommand{\thetable}{A\arabic{table}}

\begin{table*}
\caption{\label{tab01}Parameters of the various tested models resulting from the fine-time spectral analysis of GRB~131014A with their uncertainties.}
\begin{center}
{\tiny
\begin{tabular}{|c|c|c|c|c|c|c|c|c|c|c|c|c|c|c|c|c|}
\hline
\multicolumn{2}{|c|}{Time from T$_\mathrm{0}$} & \multicolumn{1}{|c|}{Models} &\multicolumn{3}{|c|}{Base Component} & \multicolumn{3}{|c|}{Additional Components} & \multicolumn{1}{|c|}{Cstat/dof} \\
\hline
 \multicolumn{2}{|c|}{} &    &\multicolumn{3}{|c|}{Band or CPL or C$_\mathrm{nTh}$} & \multicolumn{1}{|c|}{BB} & \multicolumn{2}{|c|}{C$_\mathrm{Th}$} & \multicolumn{1}{|c|}{}\\
\hline
\multicolumn{1}{|c|}{T$_\mathrm{start}$} & \multicolumn{1}{|c|}{T$_\mathrm{stop}$} & \multicolumn{1}{|c|}{Parameters} & \multicolumn{1}{|c|}{E$_{\rm peak}$} & \multicolumn{1}{|c|}{$\alpha$} & \multicolumn{1}{|c|}{$\beta$} & \multicolumn{1}{|c|}{kT} & \multicolumn{1}{|c|}{kT/E$_\mathrm{peak}$} & \multicolumn{1}{|c|}{$\alpha$} & \\
\multicolumn{1}{|c|}{(s)} & \multicolumn{1}{|c|}{(s)} & \multicolumn{1}{|c|}{} & \multicolumn{1}{|c|}{(keV)} & \multicolumn{1}{|c|}{} & \multicolumn{1}{|c|}{} & \multicolumn{1}{|c|}{(keV)} & \multicolumn{1}{|c|}{(keV)} & \multicolumn{1}{|c|}{} & \\
\hline
\multicolumn{10}{|c|}{ }   \\ 
-0.20 & +0.06 & CPL & 851.3$\pm203.0$ & +0.31$\pm0.38$ & -- & -- & -- & -- & 486/444 \\

& & BB & -- & -- & -- & 150.6$\pm14.6$ & -- & -- & 491/445 \\

& & CPL & -- & -- & -- & -- & 284.8$\pm34.8$/740.5$\pm90.48$ & +0.6 (fix) & 487/445 \\
\end{tabular}
}
\end{center}
\end{table*}

\newpage

\begin{table*}
\begin{center}
{\tiny
\begin{tabular}{|c|c|c|c|c|c|c|c|c|c|c|c|c|c|c|c|c|}
\hline
\multicolumn{2}{|c|}{Time from T$_\mathrm{0}$} & \multicolumn{1}{|c|}{Models} &\multicolumn{3}{|c|}{Base Component} & \multicolumn{3}{|c|}{Additional Components} & \multicolumn{1}{|c|}{Cstat/dof} \\
\hline
 \multicolumn{2}{|c|}{} &    &\multicolumn{3}{|c|}{Band or CPL or C$_\mathrm{nTh}$} & \multicolumn{1}{|c|}{BB} & \multicolumn{2}{|c|}{C$_\mathrm{Th}$} & \multicolumn{1}{|c|}{}\\
\hline
\multicolumn{1}{|c|}{T$_\mathrm{start}$} & \multicolumn{1}{|c|}{T$_\mathrm{stop}$} & \multicolumn{1}{|c|}{Parameters} & \multicolumn{1}{|c|}{E$_{\rm peak}$} & \multicolumn{1}{|c|}{$\alpha$} & \multicolumn{1}{|c|}{$\beta$} & \multicolumn{1}{|c|}{kT} & \multicolumn{1}{|c|}{kT/E$_\mathrm{peak}$} & \multicolumn{1}{|c|}{$\alpha$} & \\
\multicolumn{1}{|c|}{(s)} & \multicolumn{1}{|c|}{(s)} & \multicolumn{1}{|c|}{} & \multicolumn{1}{|c|}{(keV)} & \multicolumn{1}{|c|}{} & \multicolumn{1}{|c|}{} & \multicolumn{1}{|c|}{(keV)} & \multicolumn{1}{|c|}{(keV)} & \multicolumn{1}{|c|}{} & \\
\hline

\multicolumn{10}{|c|}{ }   \\
+0.74 & +0.80 & CPL & 173.5$\pm15.6$ & -0.14$\pm0.15$ & -- & -- & -- & -- & 387/444 \\

& & BB & -- & -- & -- & 33.6$\pm1.2$ & -- & -- & 435/445 \\

& &CPL & -- & -- & -- & -- & 54.8$\pm2.5$/142.5$\pm6.5$ & +0.6 (fix) & 403/445 \\
\end{tabular}
}
\end{center}
\end{table*}

\newpage

\begin{table*}
\begin{center}
{\tiny
\begin{tabular}{|c|c|c|c|c|c|c|c|c|c|c|c|c|c|c|c|c|}
\hline
\multicolumn{2}{|c|}{Time from T$_\mathrm{0}$} & \multicolumn{1}{|c|}{Models} &\multicolumn{3}{|c|}{Base Component} & \multicolumn{3}{|c|}{Additional Components} & \multicolumn{1}{|c|}{Cstat/dof} \\
\hline
 \multicolumn{2}{|c|}{} &    &\multicolumn{3}{|c|}{Band or CPL or C$_\mathrm{nTh}$} & \multicolumn{1}{|c|}{BB} & \multicolumn{2}{|c|}{C$_\mathrm{Th}$} & \multicolumn{1}{|c|}{}\\
\hline
\multicolumn{1}{|c|}{T$_\mathrm{start}$} & \multicolumn{1}{|c|}{T$_\mathrm{stop}$} & \multicolumn{1}{|c|}{Parameters} & \multicolumn{1}{|c|}{E$_{\rm peak}$} & \multicolumn{1}{|c|}{$\alpha$} & \multicolumn{1}{|c|}{$\beta$} & \multicolumn{1}{|c|}{kT} & \multicolumn{1}{|c|}{kT/E$_\mathrm{peak}$} & \multicolumn{1}{|c|}{$\alpha$} & \\
\multicolumn{1}{|c|}{(s)} & \multicolumn{1}{|c|}{(s)} & \multicolumn{1}{|c|}{} & \multicolumn{1}{|c|}{(keV)} & \multicolumn{1}{|c|}{} & \multicolumn{1}{|c|}{} & \multicolumn{1}{|c|}{(keV)} & \multicolumn{1}{|c|}{(keV)} & \multicolumn{1}{|c|}{} & \\
\hline
\multicolumn{10}{|c|}{ }   \\

+1.52 & +1.54 & Band & 687.0$\pm42.5$ & -0.03$\pm0.07$ & -3.46$\pm0.55$ & -- & -- & -- & 382/443 \\

& & CPL & 722.2$\pm36.2$ & -0.08$\pm0.06$ & -- & -- & -- & -- & 385/444 \\

& & B+BB & 802.1$\pm109.0$ & -0.12$\pm0.13$ & -3.91$\pm1.07$ & 74.7$\pm26.8$ & -- & -- & 380/441 \\
\end{tabular}
}
\end{center}
\end{table*}

\newpage

\begin{table*}
\begin{center}
{\tiny
\begin{tabular}{|c|c|c|c|c|c|c|c|c|c|c|c|c|c|c|c|c|}
\hline
\multicolumn{2}{|c|}{Time from T$_\mathrm{0}$} & \multicolumn{1}{|c|}{Models} &\multicolumn{3}{|c|}{Base Component} & \multicolumn{3}{|c|}{Additional Components} & \multicolumn{1}{|c|}{Cstat/dof} \\
\hline
 \multicolumn{2}{|c|}{} &    &\multicolumn{3}{|c|}{Band or CPL or C$_\mathrm{nTh}$} & \multicolumn{1}{|c|}{BB} & \multicolumn{2}{|c|}{C$_\mathrm{Th}$} & \multicolumn{1}{|c|}{}\\
\hline
\multicolumn{1}{|c|}{T$_\mathrm{start}$} & \multicolumn{1}{|c|}{T$_\mathrm{stop}$} & \multicolumn{1}{|c|}{Parameters} & \multicolumn{1}{|c|}{E$_{\rm peak}$} & \multicolumn{1}{|c|}{$\alpha$} & \multicolumn{1}{|c|}{$\beta$} & \multicolumn{1}{|c|}{kT} & \multicolumn{1}{|c|}{kT/E$_\mathrm{peak}$} & \multicolumn{1}{|c|}{$\alpha$} & \\
\multicolumn{1}{|c|}{(s)} & \multicolumn{1}{|c|}{(s)} & \multicolumn{1}{|c|}{} & \multicolumn{1}{|c|}{(keV)} & \multicolumn{1}{|c|}{} & \multicolumn{1}{|c|}{} & \multicolumn{1}{|c|}{(keV)} & \multicolumn{1}{|c|}{(keV)} & \multicolumn{1}{|c|}{} & \\
\hline
\multicolumn{10}{|c|}{ }   \\

+1.80 & +1.82 & Band & 485.2$\pm32.5$ & -0.01$\pm0.08$ & -2.41$\pm0.12$ & -- & -- & -- & 431/443 \\

& & CPL & 649.2$\pm27.4$ & -0.23$\pm0.05$ & -- & -- & -- & -- & 474/444 \\

& & B+BB & 981.8$\pm163.0$ & -0.43$\pm0.10$ & -2.93$\pm0.35$ & 65.2$\pm6.7$ & -- & -- & 423/441 \\
\end{tabular}
}
\end{center}
\end{table*}

\newpage

\begin{table*}
\begin{center}
{\tiny
\begin{tabular}{|c|c|c|c|c|c|c|c|c|c|c|c|c|c|c|c|c|}
\hline
\multicolumn{2}{|c|}{Time from T$_\mathrm{0}$} & \multicolumn{1}{|c|}{Models} &\multicolumn{3}{|c|}{Base Component} & \multicolumn{3}{|c|}{Additional Components} & \multicolumn{1}{|c|}{Cstat/dof} \\
\hline
 \multicolumn{2}{|c|}{} &    &\multicolumn{3}{|c|}{Band or CPL or C$_\mathrm{nTh}$} & \multicolumn{1}{|c|}{BB} & \multicolumn{2}{|c|}{C$_\mathrm{Th}$} & \multicolumn{1}{|c|}{}\\
\hline
\multicolumn{1}{|c|}{T$_\mathrm{start}$} & \multicolumn{1}{|c|}{T$_\mathrm{stop}$} & \multicolumn{1}{|c|}{Parameters} & \multicolumn{1}{|c|}{E$_{\rm peak}$} & \multicolumn{1}{|c|}{$\alpha$} & \multicolumn{1}{|c|}{$\beta$} & \multicolumn{1}{|c|}{kT} & \multicolumn{1}{|c|}{kT/E$_\mathrm{peak}$} & \multicolumn{1}{|c|}{$\alpha$} & \\
\multicolumn{1}{|c|}{(s)} & \multicolumn{1}{|c|}{(s)} & \multicolumn{1}{|c|}{} & \multicolumn{1}{|c|}{(keV)} & \multicolumn{1}{|c|}{} & \multicolumn{1}{|c|}{} & \multicolumn{1}{|c|}{(keV)} & \multicolumn{1}{|c|}{(keV)} & \multicolumn{1}{|c|}{} & \\
\hline
\multicolumn{10}{|c|}{ }   \\ 

+2.18 & +2.20 & Band & 353.9$\pm17.6$ & +0.06$\pm0.08$ & -3.81$\pm0.74$ & -- & -- & -- & 419/443 \\

& & CPL & 362.6$\pm14.5$ & +0.03$\pm0.07$ & -- & -- & -- & -- & 420/444 \\

& & C+BB & 386.9$\pm23.0$ & +0.19$\pm0.21$ & -- & 26.1$\pm7.0$ & -- & --  & 416/442 \\
\end{tabular}
}
\end{center}
\end{table*}

\newpage

\begin{table*}
\begin{center}
{\tiny
\begin{tabular}{|c|c|c|c|c|c|c|c|c|c|c|c|c|c|c|c|c|}
\hline
\multicolumn{2}{|c|}{Time from T$_\mathrm{0}$} & \multicolumn{1}{|c|}{Models} &\multicolumn{3}{|c|}{Base Component} & \multicolumn{3}{|c|}{Additional Components} & \multicolumn{1}{|c|}{Cstat/dof} \\
\hline
 \multicolumn{2}{|c|}{} &    &\multicolumn{3}{|c|}{Band or CPL or C$_\mathrm{nTh}$} & \multicolumn{1}{|c|}{BB} & \multicolumn{2}{|c|}{C$_\mathrm{Th}$} & \multicolumn{1}{|c|}{}\\
\hline
\multicolumn{1}{|c|}{T$_\mathrm{start}$} & \multicolumn{1}{|c|}{T$_\mathrm{stop}$} & \multicolumn{1}{|c|}{Parameters} & \multicolumn{1}{|c|}{E$_{\rm peak}$} & \multicolumn{1}{|c|}{$\alpha$} & \multicolumn{1}{|c|}{$\beta$} & \multicolumn{1}{|c|}{kT} & \multicolumn{1}{|c|}{kT/E$_\mathrm{peak}$} & \multicolumn{1}{|c|}{$\alpha$} & \\
\multicolumn{1}{|c|}{(s)} & \multicolumn{1}{|c|}{(s)} & \multicolumn{1}{|c|}{} & \multicolumn{1}{|c|}{(keV)} & \multicolumn{1}{|c|}{} & \multicolumn{1}{|c|}{} & \multicolumn{1}{|c|}{(keV)} & \multicolumn{1}{|c|}{(keV)} & \multicolumn{1}{|c|}{} & \\
\hline

\multicolumn{10}{|c|}{ }   \\

+3.30 & +3.40 & Band & 310.4$\pm12.7$ & -0.31$\pm0.04$ & -3.16$\pm0.29$ & -- & -- & -- & 472/443 \\

& & CPL & -- & -- & -- & -- & 329.6$\pm9.6$ (E$_\mathrm{peak}$) & -0.35$\pm0.04$ & 478/444 \\

& & B+BB & 372.5$\pm37.1$ & -0.43$\pm0.10$ & -3.56$\pm0.72$ & 35.3$\pm6.3$ & -- & --  & 465/441 \\
\hline

\end{tabular}
}
\end{center}
\end{table*}

\end{document}